\documentclass[twocolumn,showpacs,preprintnumbers,amsmath,amssymb,aps,prb,longbibliography]{revtex4-2}
\usepackage{graphicx}
\usepackage{dcolumn}
\usepackage{bm}
\usepackage{verbatim}
\usepackage[usenames, dvipsnames]{color}
\usepackage{color}
\usepackage{multirow}
\usepackage{tabularx}
\usepackage{dsfont}
\begin{document}

\title{Zeeman and Davydov splitting of Frenkel excitons in the antiferromagnet CuB$_2$O$_4$}
\author{N.~E.~Kopteva,$^{1}$ D.~Kudlacik,$^{1}$ D.~R.~Yakovlev,$^{1,2}$ M.~V.~Eremin,$^{3}$ A.~R.~Nurmukhametov,$^{3}$ M.~Bayer,$^{1,2}$ and R.~V.~Pisarev$^{2}$} 

\affiliation{$^{1}$Experimentelle Physik 2, Technische Universit\"at Dortmund, 44221 Dortmund, Germany}
\affiliation{$^{2}$Ioffe Institute, Russian Academy of Sciences, 194021 St. Petersburg, Russia}
\affiliation{$^{3}$Institute of Physics, Kazan Federal University, 420008 Kazan, Russia}

\date{\today}

\begin{abstract}
The optical spectra of antiferromagnetic copper metaborate CuB$_2$O$_4$ are characterized by an exceptionally rich structure of narrow absorption lines due to electronic transitions within the magnetic Cu$^{2+}$ ions, but their unambiguous identification and behavior in  magnetic field remain far from being fully understood. We studied the polarized magneto-absorption spectra of this tetragonal antiferromagnet with a high spectral resolution in the range of $1.4055-1.4065$~eV in magnetic fields up to 9.5~T and temperatures from 1.6 up to $T_N$~=~20~K. We observed a set of eight absorption lines at $T$~=1.6~K in magnetic fields exceeding 1.4~T which we identified as arising from Frenkel excitons related to the ground and the first excited state of Cu$^{2+}$ ions.  The number of these excitons is defined by the presence of the four Cu$^{2+}$ ions with the doubly-degenerate spin state $S = 1/2$ at the 4$b$ positions in the crystallographic unit cell. The energies of these excitons are determined the exchange interaction of 0.5~meV of Cu$^{2+}$ ions in the excited state with surrounding ions and by the Davydov splitting of 0.12~meV. In large magnetic field the observed Zeeman splitting is controlled by the anisotropic $g$-factors of both the ground and excited states. We developed a theoretical model of Frenkel excitons in magnetic field that accounts for specific features of the spin structure and exchange interactions in CuB$_2$O$_4$. The model was used for fitting the experimental data and evaluation of Frenkel exciton parameters, such as the Davydov splitting, the molecular exchange energy, and the $g$-factors of the ground and excited states of the Cu$^{2+}$ ions.
\end{abstract}
\maketitle
\section{Introduction}
\label{Introduction}

The compound with the chemical formula $\mathrm{CuB}_2\mathrm{O}_4$ has been known for over a hundred years~\cite{Mendeleev:1900}, but its crystal structure has been resolved only in 1971~\cite{Ripoll1971} and later refined in 1981~\cite{Abdullaev1981}. This material has received a lot of attention during the last two decades due to its intriguing magnetic, optical, magneto-optical and nonlinear optical properties. The magnetic phase diagram of $\mathrm{CuB}_2\mathrm{O}_4$ below $T_N=20$~K~\cite{Pankrats2018} is very complicated and includes commensurate and incommensurate phases and also phases of unknown structure. The phase transitions of different types are sensitive to  temperature and  applied magnetic field~\cite{Boehm2003,Pankrats2018,Kawamata2019}.

Optical absorption spectra of $\mathrm{CuB}_2\mathrm{O}_4$ below $T_N$ are characterized by an abundance of large number of very narrow lines, some of which have been identified with purely electronic transitions of Cu$^{2+}$ ions,  while others with phonon sidebands~\cite{Pisarev2004,Pisarev2011}. Below $T_N$, both the spatial inversion and the time reversal symmetry are broken, which opens up new possibilities for observing unusual optical phenomena. Some examples are the magnetic-field-induced optical second harmonic generation~\cite{Fiebig2003,Pisarev2004}, giant optical magnetoelectric effect~\cite{Saito2008}, one-way transparency of light~\cite{Toyoda2015}, and some others~\cite{Toyoda2019}. Very unusual effect of direction dependent luminescence induced by magnetic fields has been observed in the photoluminescence of $\mathrm{CuB}_2\mathrm{O}_4$~\cite{Toyoda2016}. Note, that the efficient photoluminescence  itself observed in $\mathrm{CuB}_2\mathrm{O}_4$\cite{Toyoda2016,Kudlacik2020} is quite surprising for Cu$^{2+}$ oxide compounds. A huge resonance nonreciprocity reaching almost 100$\%$ under the reversal of an applied magnetic field has been recently observed in studies of optical second harmonic generation in the range of the lowest-in-energy electronic transition around $1.4055-1.4061$~eV~\cite{Mund2021,Toyoda2021}. Some observations related to the action of a magnetic field on optical effects, in particular those concerning optical chirality, raised a hot dispute between experimental and theoretical groups~\cite{Arima2008,ArimaC2009,Lovesey2009,LoveseyC2009}. 

Despite the significant number of experimental studies of diverse optical effects in $\mathrm{CuB}_2\mathrm{O}_4$, until now their explanation has mostly remained at the macroscopic level, where only the factors determining the crystallographic and magnetic symmetries at temperatures below $T_N$ are taken into account~\cite{Nikitchenko}. However, the origin and understanding of the observed effects at the microscopic level is still unclear since the details of the electronic structure of the excited states of the magnetic Cu$^{2+}$ ions remain unexplored. It was suggested, that some of them can be related to the Davydov splitting of excited states~\cite{Boldyrev2015} because $\mathrm{CuB}_2\mathrm{O}_4$ is an antiferromagnet in which magnetic Cu$^{2+}$ ions occupy two types of crystallographically nonequivalent positions in the unit cell~\cite{Ripoll1971}. Namely, there are four of such ions in the 4$b$ magnetic subsystem and eight ions in the 8$d$ subsystem. Thus, the narrow absorption lines observed in previous experiments serve as a solid basis for testing the Frenkel exciton concept and for developing a relevant microscopic model. 

In the papers of Frenkel~\cite{Frenkel1,Frenkel2} and Davydov~\cite{Davydov1964,Davydov1971}, the Coulomb interaction, or more precisely its expansion in multipole moments, was suggested to be responsible for the resonant excitation transfer between the equivalent ions in a crystallographic unit cell or in a single molecule. The concept of Frenkel excitons in the optical spectra of insulating antiferromagnets was first suggested in a review paper~\cite{Loudon1968}, but it could only be applied to a few materials where the absorption lines are sufficiently narrow to allow the study of the magnetic field effects on the exciton states. 

The most exemplary materials are Cr$^{3+}$-based antiferromagnets, such as Cr$_2$O$_3$~\cite{Ziel1967} and rare earth ($R$) orthochromites $R$CrO$_3$~\cite{Aoyagi1969,Meltzer1968}. Their optical absorption spectra have very narrow lines with widths of hundred meV, which is unusual for most of antiferromagnetic oxides. Davydov splittings comparable with the linewidth have been reported. However, later the interpretation  of the Davydov splittings given in these papers was criticized in~Ref.~[\onlinecite{Allen1969}], where the Davydov splitting of excitons in Cr$_2$O$_3$ was studied experimentally and theoretically in great details. It was shown, that the previous interpretations conflict with the results of a group-theory analysis concerning interionic exchange interactions and the effect of an applied magnetic field, and a new assignment of the exciton lines was suggested. 
Soon the detailed theoretical analysis was extended to the excitons in YCrO$_3$, which showed that previous conclusions on the exciton structure should be reinterpreted~\cite{ALLEN1970,Meltzer1970,Sugano1971}.

The problem was that the magnetic field effects, which prove the involvement of the spin degrees of freedom in the formation of band excitations, were observed only on the four lines in the absorption spectra of Cr$^{3+}$ based antiferromagnets. However, taking into account the general idea about the number of the Davydov-split Frenkel exciton states, the number of such states should be larger. Indeed, there are four Cr$^{3+}$ ions in the unit cell of Cr$_2$O$_3$ and $R$CrO$_3$ crystals. In this case, each of the excited $^2$E states of the Cr$^{3+}$ ions is twofold degenerate in the orbital and twice in the spin ($S_z=\pm1/2$) variables. Thus, sixteen exciton states at $k=0$ are expected to be present, but only four lines were resolved in the absorption spectra of Cr$_2$O$_3$ and $R$CrO$_3$ crystals. We are not aware of any publication of experimental results on the observation of the whole set of Davydov-split Frenkel exciton states.  

The goal of our present paper is to study experimentally and explain theoretically on a microscopical level the fine structure of the Davydov-split Frenkel excitons in $\mathrm{CuB}_2\mathrm{O}_4$ in the $1.4055-1.4065$~eV spectral range of the electronic transitions in the 4$b$ subsystem of Cu$^{2+}$ ions. Similar to the Cr$^{3+}$-based antiferromagnets discussed above, the unit cell of $\mathrm{CuB}_2\mathrm{O}_4$ contains four magnetic Cu$^{2+}$ ions at the 4$b$ positions. However, in contrast to chromium compounds in which the spin $S$ = 3/2 in the ground state, and $S = 1/2$ in the excited state, in $\mathrm{CuB}_2\mathrm{O}_4$ the spin $S= 1/2$ is both the ground $|\epsilon\rangle = |x^2-y^2\rangle$ and excited $|\zeta\rangle = |xy\rangle$ states. Thus, both the ground and excited states of the Cu$^{2+}$ 4$b$ ions are orbitally nondegenerate and the maximum number of exciton states should be eight,   when the two spin ($S_z=\pm1/2$) states are taken into account. We succeeded to measure these eight exciton states in absorption and trace their behavior in external magnetic fields up to 9.5~T in a temperature range of $1.6-20$~K.  To the best of our knowledge, this is the first observation of the full set of the Davydov-split Frenkel  exciton states in antiferromagnets. We developed a microscopic theory, which explains well the experimental observations of Davydov-split Frenkel excitons in $\mathrm{CuB}_2\mathrm{O}_4$.

The paper is organized as follows. In Sec.~\ref{sec:2}, we discuss the crystal structure, magnetic and optical properties of CuB$_2$O$_4$ on the basis of previous studies. In Sec.~\ref{SecII}, the experimental details are given. In Sec.~\ref{Sec_Results}, we present experimental results on the absorption spectra of the lowest energy electronic transition in strong magnetic fields at cryogenic temperatures. In Sec.~\ref{Sec_Theory}, a detailed theoretical analysis of the Davydov-split Frenkel exciton states is presented. Sec.~\ref{modeling} is devoted to the modelling of the experimental results on the basis of the developed theory on the properties of the Davydov-split Frenkel exciton states in a magnetic field applied along the main crystallographic axes.
In Sec.~\ref{Conclusions}, we summarize in brief on the experimental and theoretical results.

\section{\label{sec:2}  Crystal structure, magnetic and optical properties of C\lowercase{u}B$_2$O$_4$ }

The crystallographic unit cell of $\mathrm{CuB}_2\mathrm{O}_4$ contains twelve formula units and the crystal structure is described by the tetragonal noncentrosymetric point group $\bar{4}2m$ ($D_{2d}$) and the space group $I\bar42d$ ($D_{2d}^{12}$)~\cite{Ripoll1971,Abdullaev1981}. The unit cell contains twelve magnetic copper ions  Cu$^{2+}$ (3$d^9$ electronic shell, $S=1/2$),  which occupy the 4$b$ and 8$d$ crystallographically different positions with the site symmetry $\bar4$ and $2$, respectively.  Above $T_N=20$~K~\cite{Pankrats2018}, $\mathrm{CuB}_2\mathrm{O}_4$ is in the paramagnetic phase and both the $4b$ and $8d$ magnetic subsystems of Cu$^{2+}$ ions  are disordered.  Below $T_N$, the 4$b$ subsystem is antiferromagnetically ordered, while the 8$d$ subsystem is in a partially ordered state down to the lowest temperatures~\cite{Boehm2003}. We restrict our experimental studies and theoretical analysis to the optical absorption in the spectral range of $1.4055-1.4061$~eV related to the lowest exciton transitions in the 4$b$ magnetic Cu$^{2+}$ subsystem.

\begin{figure}[t!]
\begin{center}
\includegraphics[width=8.6cm]{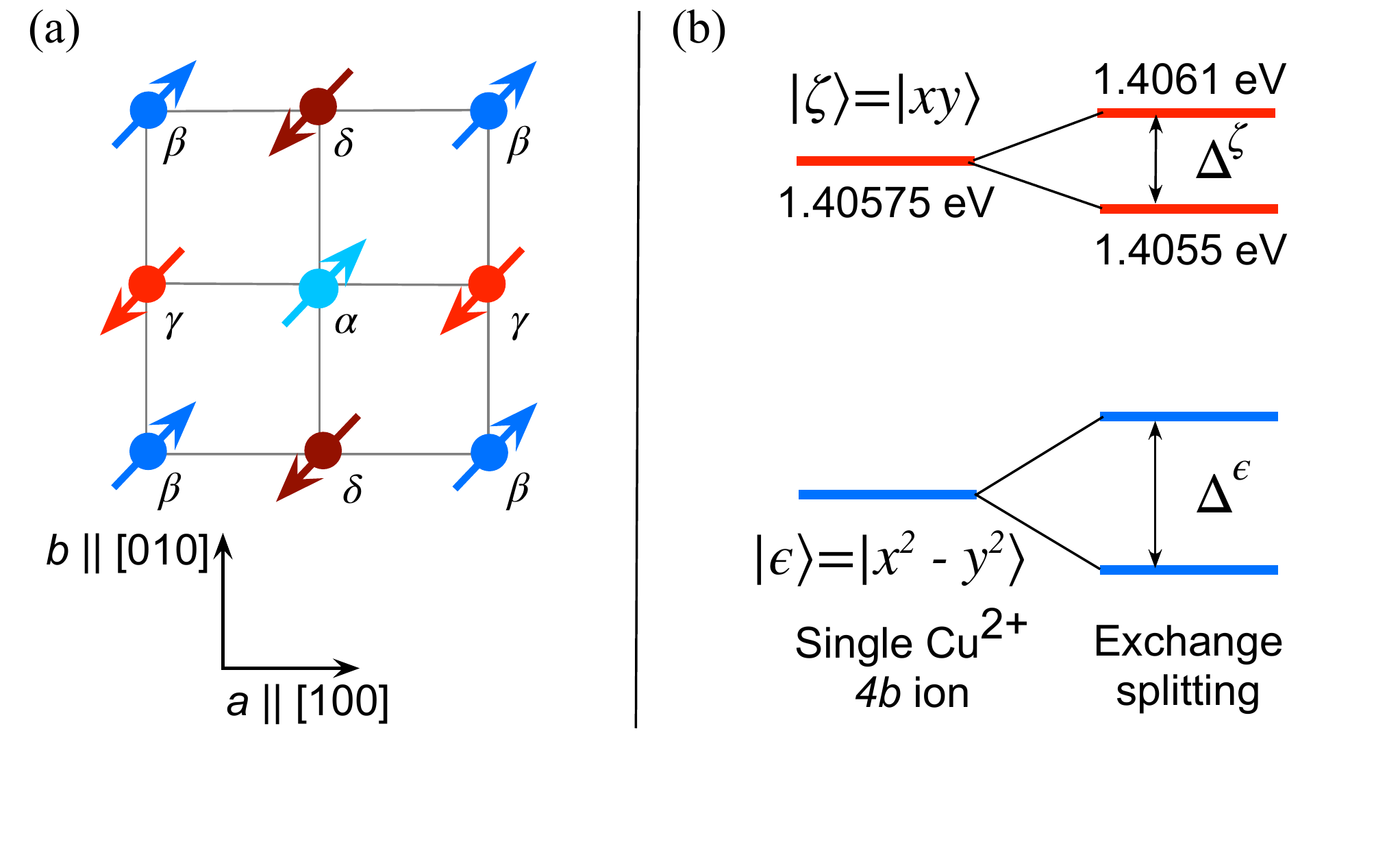}
\caption{\label{fig:Magn:Str} (a) Projection along the tetragonal $c$ axis of the antiferromagnetic structure of $\mathrm{CuB}_2\mathrm{O}_4$ for the 4$b$ magnetic subsystem in the commensurate phase in zero magnetic field. Here, the $c$ axis is pointing along the $[001]$ crystallographic direction and the $ab$ basal plane is formed by the $a\parallel[100]$ and $b\parallel[010]$ crystallographic directions. Only a single antiferromagnetic domain is shown in which the spins $S_i$ are oriented along the [110] axis, whereas in other domains spins are oriented along the [$\bar1$10], [$\bar1\bar1$0] and [1$\bar1$0]  axes. (b) Energy level diagram of the ground $|\epsilon\rangle = |x^2-y^2\rangle$ and the first excited $|\zeta\rangle = |xy\rangle$ state of a single 4$b$ Cu$^{2+}$ ion, which are split by exchange interaction $\Delta^{\epsilon (\zeta)}$ with neighboring Cu$^{2+}$ ions from the 4$b$ and 8$d$ (not shown) subsystems. The splittings $\Delta^{\epsilon}$ and $\Delta^{\zeta}$ are different in the ground and excited  states as it will be shown below in Sec.~\ref{Sec_Theory}.}
\end{center}
\end{figure}

Figure~\ref{fig:Magn:Str}(a) shows the spin structure of the 4$b$ Cu$^{2+}$ subsystem in the commensurate antiferromagnetic phase where the spins are oriented along the [110] axis in zero magnetic field. In the basis of the four 4$b$ spins $\mathbf{S}_i$ ($i = {\alpha}, {\beta}, {\gamma}$, and ${\delta)}$ in the unit cell, the two ferromagnetic $\textbf{M}$  and antiferomagnetic $\textbf{L}$ order parameters in the commensurate phase can be written as $\mathbf{M} = \mathbf{S}_{\alpha} + \mathbf{S}_{\beta} + \mathbf{S}_{\gamma} + \mathbf{S}_{\delta}$ and $\mathbf{L} = \mathbf{S}_{\alpha} - \mathbf{S}_{\beta} + \mathbf{S}_{\gamma} - \mathbf{S}_{\delta}$~\cite{Boehm2003}. Figure~\ref{fig:Magn:Str}(a) shows the Cu$^{2+}$ ions placed in different $ab$ layers along the $c$ axis in which spins are marked in light blue ($\alpha$), blue ($\beta$), red ($\gamma$) and brown ($\delta$) colors. The arrows represent the Cu$^{2+}$ spin directions along the [110] axis in the ground antiferromagnetic state without external magnetic field.

Previous studies showed that optical absorption of $\mathrm{CuB}_2\mathrm{O}_4$ in the near infrared and visible spectral range from 1.4 up to 2.5~eV arises due to electronic transitions within the 3$d^9$ electronic states of the Cu$^{2+}$ ions, which are  split by the crystal field at the two crystallographic positions~\cite{Pisarev2004,Pisarev2011}. These transitions provide exceptionally rich and highly polarized spectra with absorption coefficients reaching values above 600~cm$^{-1}$. The transparency window between 2.5 and 3.5~eV defines the blue color of CuB$_2$O$_4$ single crystals~\cite{Pisarev2011}. According to recent ellipsometric studies, strong absorption with coefficients reaching $10^6$~cm$^{-1}$ due to parity-allowed charge-transfer transitions begins at about ~4.0~eV~\cite{Rea:2021}.
 
In this paper we focus solely on the lowest photon energy electronic transition of Cu$^{2+}$ ions in the range of $1.4055-1.4061$~eV  whose energy level diagram of the ground and excited states is schematically shown in Fig.~\ref{fig:Magn:Str}(b). The electronic transition takes place between the orbitally nondegenerate ground $|\epsilon\rangle = |x^2-y^2\rangle$ and excited $|\zeta\rangle = |xy\rangle$ states of a single Cu$^{2+}$ ion \cite{Pisarev2004,Pisarev2011,Boldyrev2015}. Due to the exchange interaction of  the Cu$^{2+}$ ion with surrounding ions in the 4$b$ and 8$d$ subsystems, these states are 
split by the $\Delta^{\zeta}$ and $\Delta^{\epsilon}$ values, respectively. The main task of our paper is to show that this single-ion scheme used in several previous publications on optical experiments in this spectral region should be seriously revised due to the presence of the four 4$b$ Cu$^{2+}$ ions in the elementary crystallographic unit cell~\cite{Ripoll1971} interacting with surrounding Cu$^{2+}$ ions from both the 4$b$ and 8$d$ magnetic subsystems. The concept of Frenkel excitons and Davydov splitting is applied for explaining the experimentally observed fine structure of the absorption in an applied  magnetic field.    

\section{Experimental details}
\label{SecII}

Single crystals of CuB$_2$O$_4$ were grown by the Kyropoulos technique from a melt of B$_2$O$_3$, CuO, Li$_2$O, and MoO$_3$ oxides~\cite{Petrakovskii2000,Fiebig2003,Pisarev2004,Pisarev2011}. To ensure a well-defined orientation of the samples, a plane-parallel polished plates were cut from single crystals oriented using Laue X-ray diffraction patterns.  

\begin{figure}[b!]
\begin{center}
\includegraphics[width=7.0cm]{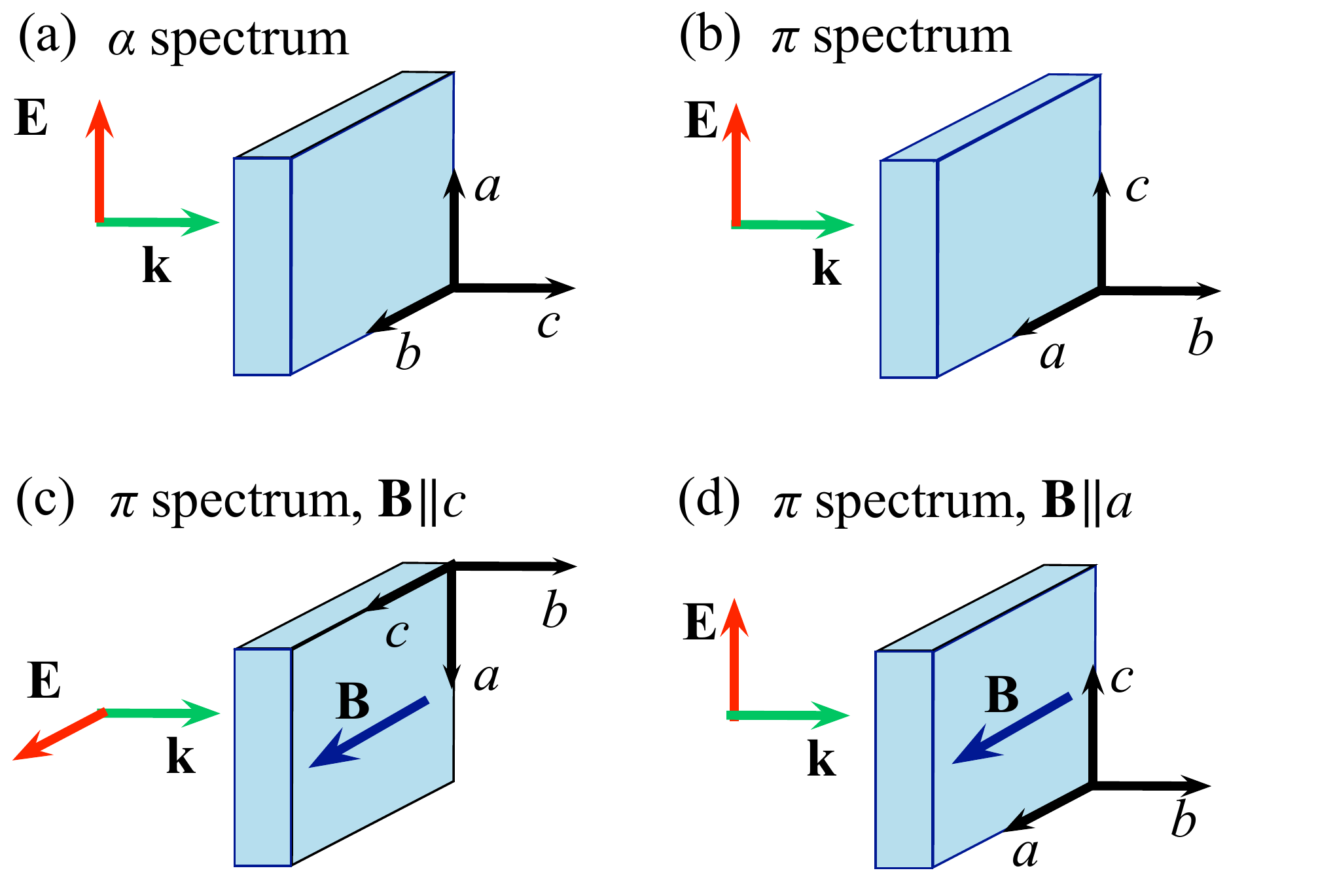}
\caption{\label{geometries} Schemes of the sample axes orientation with respect to the $\textbf{k}$-vector of light and the light polarization vector $\textbf{E}$ for the $\alpha$ (a) and $\pi$ (b) spectra. The magnetic field is applied in Voigt geometries $\textbf{k}\perp \textbf{B} \parallel c$ (c) and $\textbf{k}\perp \textbf{B} \parallel a$ (d) for the measurement of the $\pi$ spectrum.}
\end{center}
\end{figure}

Two samples were studied, namely the (001) sample with the optical $c$-axis  oriented along the sample normal (see Fig.~\ref{geometries}(a)),
$c \parallel [001]$, and the (101) sample with the optical axis $c \parallel [001]$ in the sample plane (see Fig.~\ref{geometries}(b)). The $ab$ basal plane in the (001) sample is perpendicular to the $[001]$ crystallographic direction and is formed by the two $a$ and $b$ crystallographic axes ($a \parallel [100]$ and $b \parallel [010]$).

The $1.12$~mm thick (101) sample was used for reliable recording of $\pi$ spectra with absorption on the order of 10~cm$^{-1}$ when light wave vector $\textbf{k} \parallel b$ and the polarization of the incident light $\textbf{E} \parallel c$. The (001) sample with a thickness of $59~\mu$m was used to detect the $\alpha$ spectra for the $\textbf{k} \parallel c$ and $\textbf{E} \parallel a$ geometry. 

The samples were mounted strain-free in a split-coil magnet cryostat and measurements were performed in the temperature range $1.6-20$~K. The magnetic field was oriented perpendicular to the $\textbf{k}$ vector of light (Voigt geometry) and its value was tuned from 0 up to 9.5~T.  The (101) sample was rotated in the $ac$  plane by 90$^\circ$ to obtain the $\textbf{B} \parallel c$ and $\textbf{B} \parallel a$ geometry, see Figs.~\ref{geometries}(c) and \ref{geometries}(d) respectively.

For investigating the absorption spectra in the spectral range from 1.35 to 2.5~eV, the white light from a halogen lamp was used. Power density of the light source was set to $5$~$\mu$W/mm$^2$. Linearly polarized light was set by a Glan-Thompson prism. A $\lambda/2$ retardation plate was used for rotating the linear polarization between  vertical and horizontal orientations.  In order to avoid the polarization characteristics of the grating spectrometer, we send through it the light of fixed linear polarization. For that the light transmitted through the sample was analyzed using a $\lambda/4$ retardation plate with its optical axis oriented at 45$^\circ$, and additionally a Glan-Thompson prism. 
The transmitted light was spectrally resolved by a 1~m Spex spectrometer equipped with a 10 $\times$ 10~cm$^2$ sized grating with 1200 grooves/mm used in the first order. A silicon charge-coupled device (CCD) camera with 512$\times$2048 pixels of 13.5~$\mu$m size was used as a detector. In combination with $4\times$ magnification optics in front of the CCD camera, it was possible to improve the spectral resolution to 20~$\mu$eV.

\section{Experimental results}
\label{Sec_Results}

\subsection{Absorption spectra in zero magnetic field }
\label{Sec_Abs_zero}

\begin{figure}[t]
\begin{center}
\includegraphics[width=8.6cm]{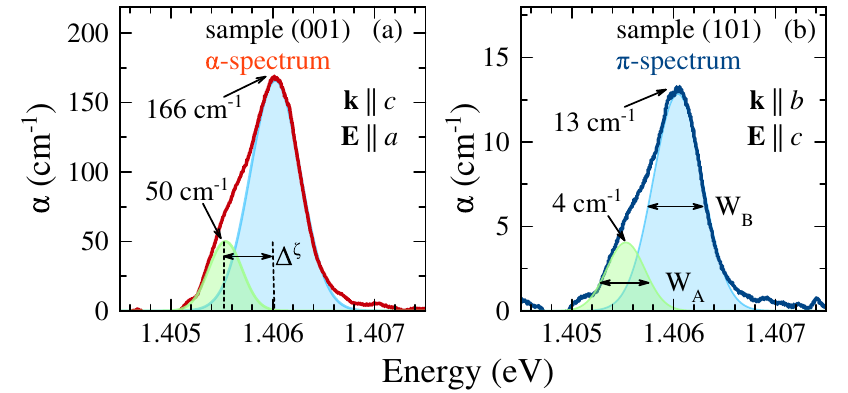}
\caption{\label{fig1}  
Polarized absorption spectra of CuB$_2$O$_4$ in the spectral range of the lowest photon-energy transition from the ground $|\epsilon\rangle = |x^2-y^2\rangle$ state to the lowest-in-energy excited $|\zeta\rangle = |xy\rangle$ state of the Cu$^{2+}$ ions at the 4$b$ positions~\cite{Pisarev2004,Pisarev2011,Boldyrev2015}. The spectra were measured at $T~=~1.6$~K in zero magnetic field for the two geometries corresponding to the $\alpha$-spectra (a) when $\textbf{k} \parallel c$ and $\textbf{E} \parallel a$ measured on the (001) sample, and the $\pi$-spectra (b) when $\textbf{k} \parallel b$ and $\textbf{E} \parallel c$ measured on the (101) sample. The absorption coefficients in the two spectra differ by more than an order of magnitude. According to the selection rules for absorption in uniaxial crystals, the $\alpha$ and $\pi$ spectra correspond to the electric-dipole (ED) and magnetic-dipole (MD) transitions, respectively~\cite{MCCLUREbook}. The green and blue shaded areas represent Gaussian fits used for the evaluation of involved exciton parameters. The FWHMs are indicated by $W_\text{A(B)}$. The doublet splitting assigned to the excited state is denoted by $\Delta^\zeta$.} 
\end{center}
\end{figure}

Low-temperature absorption spectra of CuB$_2$O$_4$ in the range of $1.4-2.5$~eV in three main polarizations ($\alpha$, $\sigma$ and $\pi$) are characterized by a rich structure of zero-phonon (ZP) lines related to the Cu$^{2+}$ ions in the 4$b$ and 8$d$ sublattices~\cite{Pisarev2004,Pisarev2011,Boldyrev2015}. Each ZP line is accompanied by a long tail of phonon sidebands. Here, we focus exclusively on the set of absorption lines originating from the first ZP line related to the 4$b$ subsystem within the $1.4055-1.4061$~eV spectral range. The $\alpha$ and $\pi$ spectra measured in the (001) and (101) samples are shown in Figs.~\ref{fig1}(a) and \ref{fig1}(b), respectively.

The observed asymmetric line originates from the optical transition between the ground $|\epsilon\rangle = |x^2-y^2\rangle$ and excited $|\zeta\rangle = |xy\rangle$ states of the Cu$^{2+}$ ions \cite{Pisarev2004,Pisarev2011,Boldyrev2015}. The asymmetric shape suggests the presence of at least two partially overlapping lines with different amplitudes, which parameters we evaluate from the double Gaussian fits shown by the green and blue areas in Fig.~\ref{fig1}. We refer to the lines at lower 1.4055~eV and higher 1.4061~eV photon energy as A and B lines, respectively. The intensity ratio between the lines is $I_\text{B}/I_\text{A} = 3.32$ and 3.25  for the $\sigma$- and $\pi$-spectra, respectively. Energy splitting between the A and B lines amounts to $\Delta^\zeta=0.5$~meV. Both lines have similar full widths at half maximum (FWHM) of $W_\text{A}=0.44$~meV and $W_\text{B}=0.56$~meV. These lines can be identified as originating from Frenkel excitons. 

It is worth noting here that this zero field splitting of the ZP absorption line, as well as the luminescence line, was previously noted in Refs.~\cite{Boldyrev2015,Kudlacik2020}, respectively. The observed ZP line splitting is puzzling because the relevant electronic transition takes place between the  ground and excited states of a single Cu$^{2+}$ ion which are both orbitally nondegenerate. 
A tentative assignment of the doublet structure of the corresponding line to the Davydov splitting was suggested due to the presence of the two copper Cu$^{2+}$ ions in the primitive unit cell~\cite{Boldyrev2015}. 
In our paper on the basis of experiments at low temperature in high magnetic field, being supplemented by a microscopic theoretical analysis, we present a relevant microscopic model of the Davydov splitting of Frenkel excitons in CuB$_2$O$_4$.

To conclude this subsection, we would like to make the following comment.
Both the ground and excited electronic states of Cu$^{2+}$ ions do not interact with the electric field of the light wave because they have the even parity of their 3$d^9$ wave functions~\cite{Eremin2019_2}. Therefore, the relevant transitions are forbidden in the electric dipole (ED) approximation. However, the local crystal field acting on these Cu$^{2+}$ ions leads to a mixing of the even $|\epsilon\rangle = |x^2-y^2\rangle$ ground state with the odd 3$d^{8}$4$p^{1}$ and charge transfer excited configurations. Due to the odd crystal field components, the mixing between these opposite parity configurations  and  oxygen-copper electron transfer process make optical transitions allowed both in the ED and MD approximations~\cite{abragam2012electron,SuganoTanabeKamimura}.

\subsection{Magneto-absorption spectra of Frenkel excitons at $\textbf{$T$=1.6}$~K}
\label{Sec_Abs_B}

\begin{figure*}[t!]
\begin{center}
\includegraphics[width=17cm]{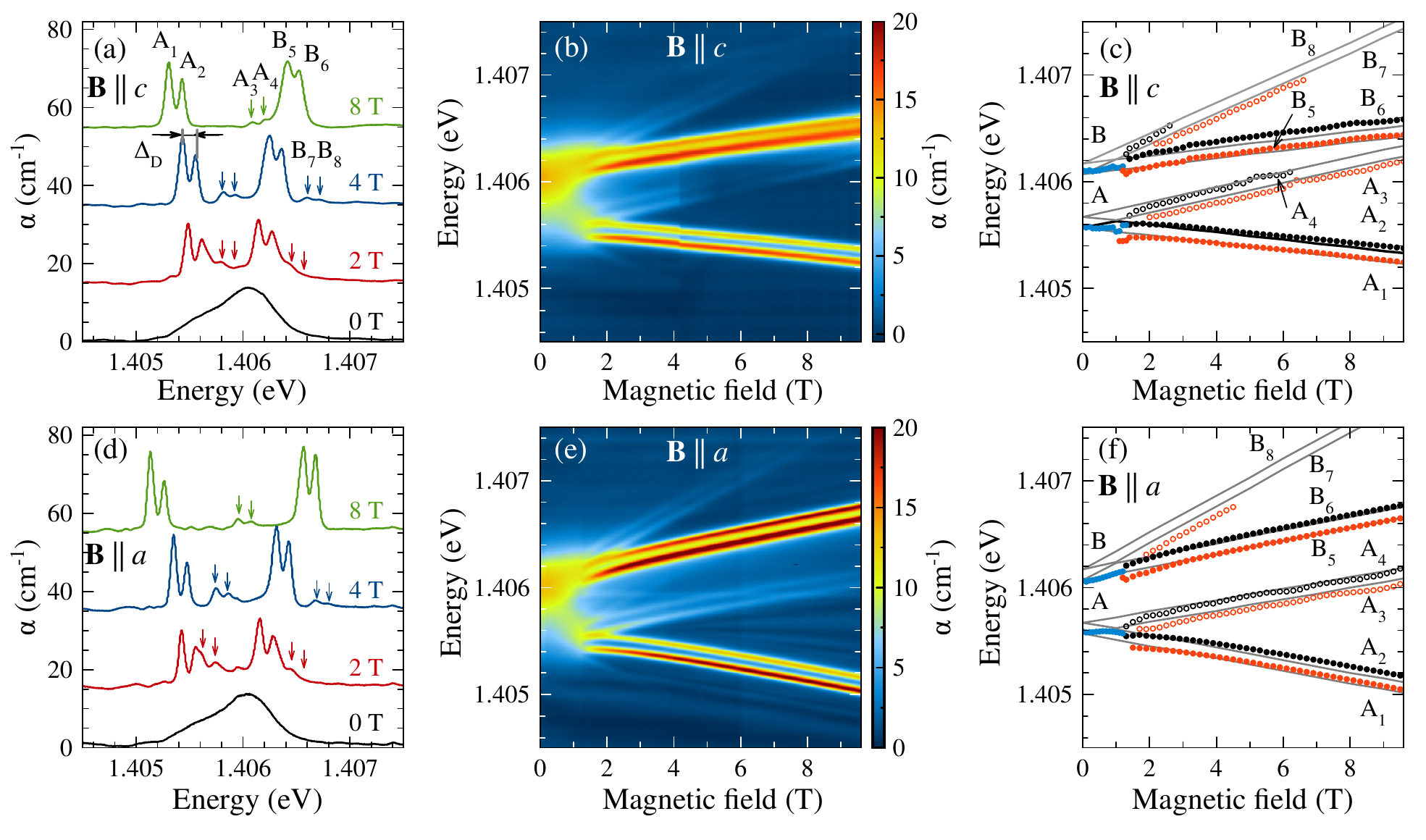}
\caption{\label{fig2}  
Evolution of the absorption spectra showing the Frenkel excitons due to optical transitions between the ground and excited states of Cu$^{2+}$ ions within the 4$b$ sublattice. The $\pi$ spectra in all panels are measured at $T$~=~1.6~K for the (101) sample. Results in panels (a), (b) and (c) are recorded in the $\textbf{B} \parallel c$ configuration. Results in panels (d), (e) and (f) are recorded in the $\textbf{B} \parallel a$ configuration. Panels (b) and (e) are contour plots where the absorption amplitude is coded with color. The magnetic field dependence of the absorption lines maxima is plotted by symbols is panels (c) and (f).  The solid lines show modeling results for the Frenkel exciton states using the theoretical approach of Sec.~\ref{Sec_Theory}. Experimental data were fitted with the following parameters:  $g_\text{$c$}^\zeta = 1.81$,  $g_\text{$c$}^\epsilon = 1.25$, $\delta g_c = 0.30$ for the (c) panel, and $g_\text{$a$}^\zeta = 1.93$, $g_\text{$a$}^\epsilon = 2.06$, $\delta g_\text{$a$} = -0.04$ for the (f) panel. Parameters $|t_\text{F}|=0.06$~meV and $\Delta^{\zeta} = 0.5$~meV are the same for the both panels.}
\end{center}
\end{figure*}

Application of the magnetic field is a very powerful tool in the exciton spectroscopy which allows one to disclose fine structure of electronic levels and make unique identification of the exciton states using their Zeeman splittings and polarization properties. In magnetically ordered  materials an applied magnetic field can induce phase transitions which can also affect the exciton states. 

We study modification of the exciton absorption spectra in $\pi$ geometry (see Fig.~2) in magnetic fields up to 9.5~T at $T=1.6$~K. Magneto-absorption results for the (101) sample are collected in Fig.~\ref{fig2} for two magnetic field orientations along the main $a$ and $c$ crystallographic axes. Results are presented in a comprehensive way by showing absorption spectra, their contour maps and field shifts of the line maxima. 

Let us first consider the $\textbf{B} \parallel c$ configuration presented in the upper row of panels in this Figure. One can see in Fig.~\ref{fig2}(a) that the application of the magnetic field drastically modifies the absorption spectra. In the field range exceeding $B_0=1.4$~T the linewidth is strongly reduced from about 0.5~meV down to about $0.06-0.14$~meV. A fan of narrow lines becomes well resolved and one can distinguish up to eight lines. This number is just the one that can be expected for the Davydov-split Frenkel excitons of the four 4$b$ Cu$^{2+}$ ions with the spin value $S$=1/2 within the unit cell, as it is discussed in detail in Sec.~\ref{Sec_Theory}. The field evolution of the spectra can be clearly observed in the contour plot in Fig.~\ref{fig2}(b). Here, a rather drastic transformation of the spectra above $B_0=1.4$~T is observed.
Below this value, the unresolved smeared absorption is observed.
In strong fields, the eight exciton lines grouped into four pairs with the same $\Delta_\text{D}=0.12$~meV splitting within each pair are reliably resolved. Independence of the splitting from of the field strength allows us to assign these lines to the Davydov-split pairs. 

\begin{figure*}[t!]
\begin{center}
\includegraphics[width=18cm]{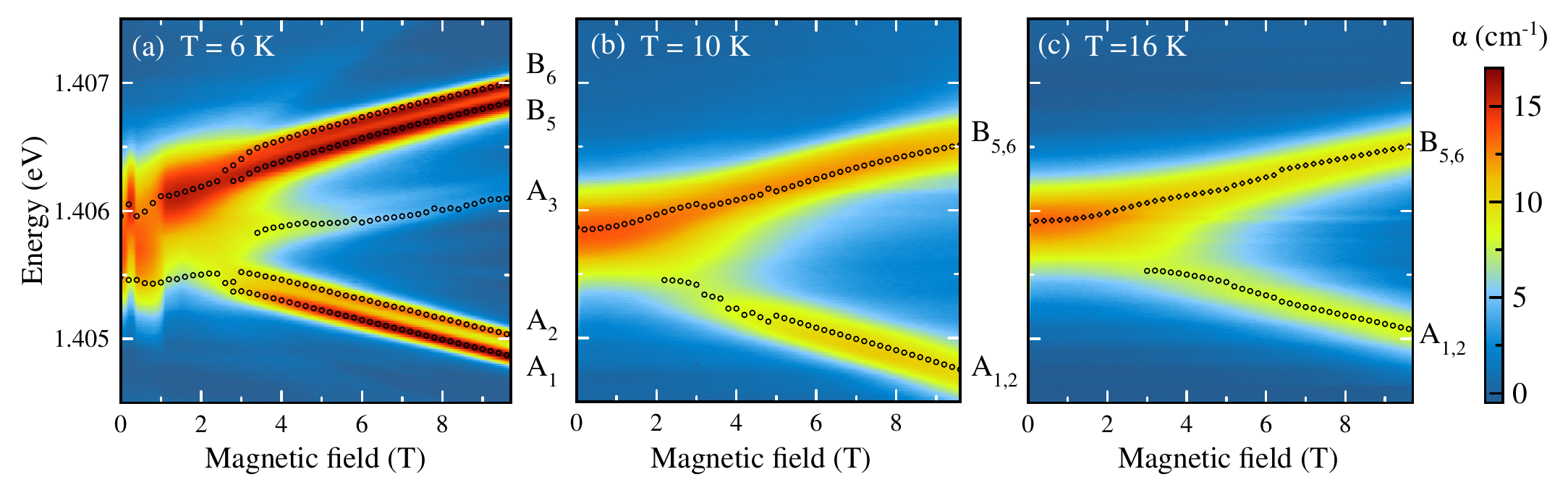}
\caption{\label{fig3}  Contour plots of exciton absorption spectra at $T=6$~K (a), 10~K (b) and  16~K (c) for the (101) sample. The $\pi$ absorption spectra are shown 
in the configuration $\textbf{E} \parallel c$, $\textbf{k} \parallel b$ and $\textbf{B} \parallel c$. Black symbols show the energies of the line maxima.} 
\end{center}
\end{figure*}

In the fields below $B_0=1.4$~T, the behavior of lines is complicated and no fine structure can be resolved presumably because of the broadening and overlapping of several lines due to the presence of antiferromagnetic domains and not well-defined spin structure~\cite{Pankrats2018}. One can see in Fig.~\ref{fig2}(c) that the four lowest-in-energy lines are originating from the A line observed in zero field (see Fig.~3) which we label from A$_1$ to A$_4$, and the four upper lines are assigned to the zero-field B line labeled from B$_5$ to B$_8$. Important to note, that two pairs of lines A$_{1,2}$ and B$_{5,6}$ have notably larger intensity than the two others. For instance, at $B=5$~T the absorption coefficient of the stronger A$_{1,2}$ lines amounts to 20~cm$^{-1}$, whereas for the A$_{3,4}$ lines it is 5~cm$^{-1}$.  One can see from the bottom row of panels in Fig.~\ref{fig2}, that in the $\textbf{B} \parallel a$ configuration the behavior is qualitatively similar. The critical field is also $B_0=1.4$~T, but the slopes of the line shifts are notably different. The latter appearance is expected as the magnetic structure is anisotropic with the antiferromagnetic spins lying predominantly in the ($ab$) plane~\cite{Pankrats2018}. Understanding of the line structure and field shifts calls for a theory which is presented in Sec.~\ref{Sec_Theory}.

Electronic excitations related to the ZP lines in 3$d^n$ insulators should most adequately be regarded as Frenkel excitons as it was discussed in earlier literature~\cite{Ziel1967,Loudon1968,Meltzer1968,Aoyagi1969,Allen1969,Meltzer1970,Sugano1971,Eremenko1970,Eremenkobook,Imbusch1978}. However, the concept of Frenkel excitons can actually be applied only to those cases when the absorption lines are narrow enough and can be well resolved in an applied magnetic field and at least at low temperature. The narrowness of the ZP absorption lines in CuB$_2$O$_4$ at $T$~=~1.6~K is just the right scenario for applying this concept.  The observation of a zero-field splitting of the absorption doublet line at low temperature and a fine splitting in applied magnetic field are very important features for the development of a theoretical model of Frenkel excitons and their Davydov splitting in CuB$_2$O$_4$ presented in detail in Sec.~\ref{Sec_Theory}.

\subsection{Temperature dependence of Frenkel excitons}
\label{Sec_Abs_T}

An increase in temperature leads to a broadening of the exciton absorption lines and modifications in 
their splittings  and magnetic field dependencies. Figure~\ref{fig3} shows the contour plots of exciton $\pi$ absorption spectra at $T=6$~K (a), 10~K (b) and  16~K (c) ($\textbf{E} \parallel c$, $\textbf{k} \parallel b$) in the $\textbf{B} \parallel c$ configuration  for the (101) sample. The spectral lines become broader and the separation between the eight exciton levels becomes complicated, but doublet structure remains resolved at $T=6$~K. One can see, that the critical field at which the spin splitting of Frenkel excitons becomes resolved is increased to $B_0 = 3$~T. For the fields above this value the fitting procedure still makes it possible to distinguish two doublets (A$_{1,2}$ and B$_{5,6}$) and the unresolved weak line A$_{3,4}$, which maxima are shown by the black symbols in Fig.~\ref{fig3}(a). Further temperature increase leads to smearing of the fine structure and only broad spectral lines are resolved at 10~K [Fig.~\ref{fig3}(b)] and at 16~K [Fig.~\ref{fig3}(c)].

\begin{table}[b!]
\caption{Slopes $\beta$ of the $A_{1-4}$ and $B_{5-8}$ groups of exciton lines evaluated from the data in magnetic field for the $\textbf{B} \parallel c$ and $\textbf{B} \parallel a$ configurations measured at different temperatures.}
\label{tab:t1}
\begin{center}
\begin{tabular*}{0.36\textwidth}{@{\extracolsep{\fill}} |>{\centering\arraybackslash} m{0.06\textwidth}|>{\centering\arraybackslash} m{0.06\textwidth}|>{\centering\arraybackslash} m{0.06\textwidth}|>{\centering\arraybackslash} m{0.06\textwidth}|>{\centering\arraybackslash} m{0.06\textwidth}|}
\hline
$T$~(K) &   A$_{1}$, A$_{2}$   & A$_{3}$, A$_{4}$ & B$_{5}$, B$_{6}$ & B$_{7}$, B$_{8}$\\
\hline
\multicolumn{5}{|c|}{$\textbf{B} \parallel c$} \\
\hline
1.6  & $-0.54$   &   1.30   &   0.65  &   2.41\\
\hline
6.0  & $-0.78$   &   1.24   &   1.08  &   \\
\hline
10.0  & $-0.99$   &      &   0.88  &   \\
\hline
16.0  &  $-0.86$   &      &   0.72  &   \\
\hline
\multicolumn{5}{|c|}{$\textbf{B} \parallel a$}\\
\hline
1.6 & $-0.94$  &   0.92   &   1.07   &   2.98\\
\hline
6.0  & $-1.32$   &   0.77   &   1.41  &   \\
\hline
10.0  & $-1.59$   &      &   1.29  &   \\
\hline
16.0  & $-1.34$   &     &   1.22  &   \\
\hline
\end{tabular*}
\end{center}
\end{table}

\begin{figure}[t!]
\begin{center}
\includegraphics[width=8.6cm]{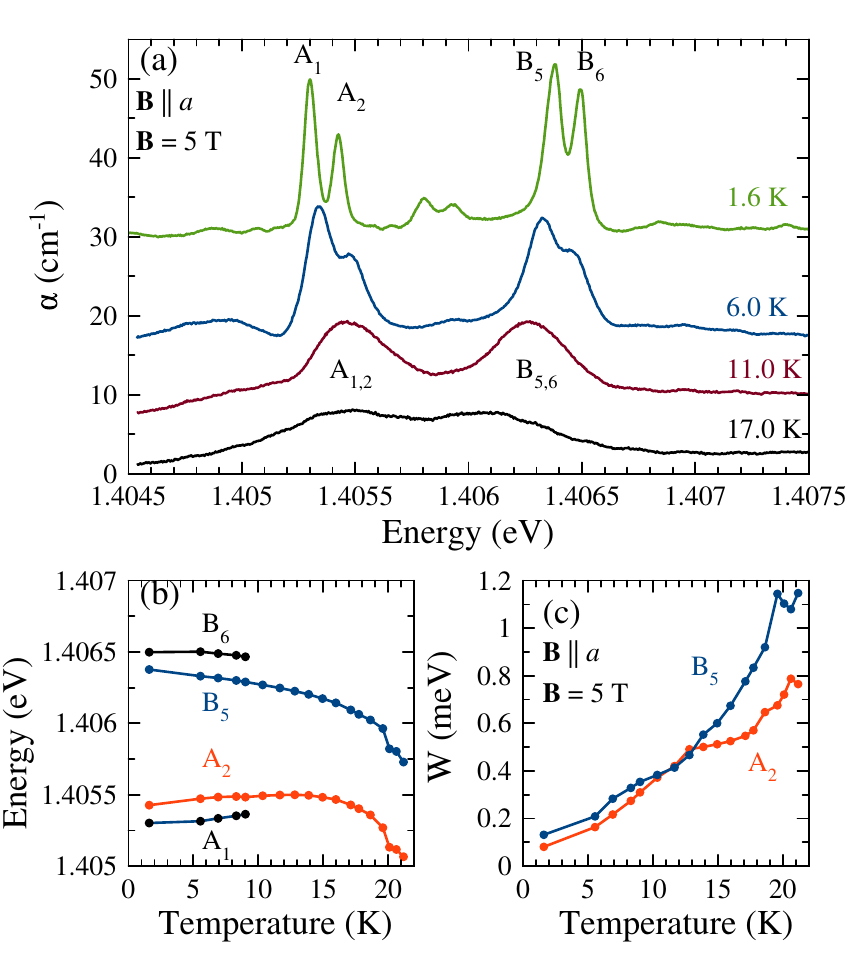}
\caption{\label{fig4} (a) $\pi$-absorption spectra of CuB$_2$O$_4$ at $B~=~5$~T in $\textbf{B}\parallel a$ geometry measured at different temperatures for the (101) sample. 
(b) Spectral positions of the A$_{1}$, A$_{2}$, B$_{5}$ and B$_{6}$ exciton lines in dependence on temperature. (c) The FWHM of the exciton lines with large amplitude in dependence on temperature.}
\end{center}
\end{figure}

A similar behavior is observed for the $\textbf{B}\parallel a$ configuration. To illustrate this, we show in Fig.~\ref{fig4}(a) an example of absorption spectra at $B_a$~=~5~T measured at various temperatures. The spectrum at $T = 1.6$~K consists of four doublets. Increasing the temperature to $T = 6$~K leads to line broadening and at $T > 9$~K the doublets are not resolved and only two broad lines are distinguishable. 

As it was shown above, the exciton fine structure depends on the mutual orientation of the external magnetic field with respect to the crystal $c$ axis. To obtain the parameters for all spectral lines shown in Figs.~\ref{fig2} and \ref{fig3}, each absorption spectrum at high magnetic fields ($B>1.4$~T) was fitted by eight Lorentzian functions for $T = 1.6$~K and 6~K but only two Lorentzian functions are used for $T > 9$~K. The spectral positions of A$_{1}$, A$_{2}$, B$_{5}$  and B$_{6}$ lines are shown in Fig.~\ref{fig4}(b) as a function of temperature. The width of the exciton lines increases from 0.08~meV up to 0.8~meV for the A$_{1}$, A$_{2}$ doublet and from 0.13~meV up to 1.1~meV for the B$_{5}$, B$_{6}$ doublet with a temperature increase up to 21~K, see Fig.~\ref{fig4}(c). 

The energy shifts of the exciton lines in magnetic field are interpolated by a linear function in order to define the slopes $\beta$ and the energy offsets $\mathds{E}_0$ at $B = 0$:

\begin{equation}
\mathds{E}(B) = \beta \mu_\text{B} B + \mathds{E}_0,     
\end{equation}
where $\mu_\text{B}$ is the Bohr magneton. 

The slopes $\beta$ of exciton lines for different field configurations and temperatures are given in Table~\ref{tab:t1}. The A and B groups consist of the two sets of doublets split by the energy $\Delta_{\text{D}}$ which does not depend on the magnetic field. The energy offsets $\mathds{E}_0$ for different temperatures are collected in Table~\ref{tab:offsets}.

\begin{table}[b!]
\caption{Zero field energy offsets (eV) at different temperatures for the $\mathds{E}_0$ of the $A_{1-4}$ and $B_{5-8}$ exciton lines. The data are evaluated from the magnetic field dependences for the $\textbf{B} \parallel c$ configuration.}
\begin{center}
\begin{tabular*}{0.38\textwidth}{@{\extracolsep{\fill}} |>{\centering\arraybackslash} m{0.06\textwidth}|>{\centering\arraybackslash} m{0.06\textwidth}|>{\centering\arraybackslash} m{0.06\textwidth}|>{\centering\arraybackslash} m{0.06\textwidth}|>{\centering\arraybackslash} m{0.06\textwidth}|}
\hline
$T~$(K) & A$_{1,2}$ & A$_{3,4}$ & B$_{5,6}$ & B$_{7,8}$ \\
\hline
\multicolumn{5}{|c|}{$\textbf{B} \parallel c$} \\ 
\hline
1.6 & 1.40561 & 1.40557 & 1.40616 & 1.40603 \\ 
\hline
5.0 & 1.40568 & 1.40571 & 1.40613 &  \\ 
\hline
10.0 & 1.40564 &  & 1.40577 &  \\ 
\hline
16.0 & 1.40580 &  & 1.40585 &  \\ 
\hline
\end{tabular*}
\end{center}
\label{tab:offsets}
\end{table}

Understanding the energy structure of Frenkel excitons in CuB$_2$O$_4$ at low temperatures and their behavior in applied magnetic fields calls for the detailed theoretical consideration which is developed in Sec.~\ref{Sec_Theory}. 

\section{Theoretical analysis of Frenkel excitons in C$\textbf{u}$B$_2${O}$_4$}
\label{Sec_Theory}
\subsection{The concept of Frenkel excitons in C$\textbf{u}$B$_2${O}$_4$}

As discussed in Sec.~\ref{Sec_Abs_zero}, the concept of Frenkel excitons was proven to be essential for understanding the optical spectra of antiferromagnetic insulators. In particular, the Frenkel excitons play a major role in the absorption of electronic transitions between the 3$d^n$ states. However, most of experiments were performed and theoretical models proposed for the chromium Cr$^{3+}$  antiferromagnets in which transitions between 3$d^3$ electronic states take place. These are the chromium oxides Cr$_2$O$_3$ and the rare-earth ($R$) orthochromites $R$CrO$_3$~\cite{Ziel1967,Loudon1968,Meltzer1968,Aoyagi1969,Allen1969,Meltzer1970,Sugano1971,Eremenko1970,Eremenkobook,Imbusch1978}. 

Before proceeding to the theoretical analysis of Frenkel excitons in CuB$_2$O$_4$, we would like to point out the similarities and differences between CuB$_2$O$_4$ and the Cr$^{3+}$ antiferromagnetic insulators. As for similarities, the number of magnetic ions in the unit cell of CuB$_2$O$_4$ in the 4$b$ subsystem and chromium crystals is the same, namely four. Thus, some similarities in the experimental results could be expected. However, in zero magnetic field only two lines are observed in CuB$_2$O$_4$ instead of four lines in Cr$^{3+}$ antiferromagnets. When the magnetic field is applied, four pairs of lines are resolved in CuB$_2$O$_4$ at the low temperature, that is, a total of eight exciton lines (see spectra in Sec.~\ref{Sec_Abs_B}). Similar results were obtained in the studies of second harmonic generation in CuB$_2$O$_4$~\cite{Mund2021}. In contrast, only four lines are found in Cr$^{3+}$ antiferromagnets. These and other results on the magneto-absorption in CuB$_2$O$_4$ motivate us to get deeper insight by developing a microscopic model of the Frenkel excitons in this antiferromagnetic material. The model accounts for the spin states of the Cu$^{2+}$ 4$b$ ions and their role in the Davydov splitting of Frenkel excitons. 
We use the cell perturbation method \cite{Ziel1967,Aoyagi1969,Allen1969} for calculating the Frenkel exciton energies which is performed in two steps.

The elementary unit cell of CuB$_2$O$_4$ contains four  Cu$^{2+}$ 4$b$ ions, which are denoted as $\alpha$, $\beta$, $\gamma$ and $\delta$, see Figs.~\ref{fig:Magn:Str}(a) and \ref{fig5}(a). The optical absorption in the $1.4055-1.4061$~eV spectral range is due to the transitions from the ground $|\epsilon\rangle = |x^2 - y^2\rangle$ state of the Cu$^{2+}$ ion to the excited state $|\zeta\rangle = |xy\rangle$ at 4$b$ position, see Fig.~\ref{fig:Magn:Str}(b)~\cite{Pisarev2011}. Both these states are orbital singlets which are twofold spin $S$~=~1/2 degenerate ($m_s=\pm1/2$). In the ordered antiferromagnetic phase below $T_N$, the splitting of the ground state under the action of the exchange (molecular) field caused by the neighboring copper spins is about 7.2~meV \cite{Toyoda2016}.  Therefore, it can be reasonably assumed that at low temperatures only the lower $m_s~=-1/2$ substate of the spin doublet is populated. The magnitude of the exchange splitting of the excited state is not known, and its determination is one of the tasks of our theory. 

Other important tasks to clarify are the particular features of the exciton states and the parameter of the energy transfer of resonant excitation between the four Cu$^{2+}$ ions occupying the $\alpha$, $\beta$, $\gamma$, and $\delta$ positions within the elementary unit cell. Since the wavelength corresponding to optical absorption is much larger than the lattice period, in accordance with the law of momentum conservation, exciton transitions  with only small values of the wave vector $\textbf{q}$ can be excited and probed optically. The problem of calculating $q~=~0$ states is reduced to diagonalization of the interaction energy between the $\alpha$, $\beta$, $\gamma$, and $\delta$ ions within the basis of wave functions corresponding to a single unit cell.

These wave functions can be written in the following form:
\begin{gather}
\label{EQ01}
|\psi_1\rangle = |\zeta_{\alpha}+\rangle |\epsilon_{\beta}+\rangle |\epsilon_{\gamma}-\rangle|\epsilon_{\delta}-\rangle,\notag\\ 
|\psi_2\rangle = |\zeta_{\alpha}-\rangle |\epsilon_{\beta}+\rangle |\epsilon_{\gamma}-\rangle|\epsilon_{\delta}-\rangle,\notag\\
|\psi_3\rangle = |\epsilon_{\alpha}+\rangle |\zeta_{\beta}+\rangle |\epsilon_{\gamma}-\rangle|\epsilon_{\delta}-\rangle,\notag\\
|\psi_4\rangle = |\epsilon_{\alpha}+\rangle |\zeta_{\beta}-\rangle |\epsilon_{\gamma}-\rangle|\epsilon_{\delta}-\rangle,\\
|\psi_5\rangle = |\epsilon_{\alpha}+\rangle |\epsilon_{\beta}+\rangle |\zeta_{\gamma}+\rangle|\epsilon_{\delta}-\rangle,\notag\\
|\psi_6\rangle = |\epsilon_{\alpha}+\rangle |\epsilon_{\beta}+\rangle |\zeta_{\gamma}-\rangle|\epsilon_{\delta}-\rangle,\notag\\
|\psi_7\rangle = |\epsilon_{\alpha}+\rangle |\epsilon_{\beta}+\rangle |\epsilon_{\gamma}-\rangle|\zeta_{\delta}+\rangle,\notag\\
|\psi_8\rangle = |\epsilon_{\alpha}+\rangle |\epsilon_{\beta}+\rangle |\epsilon_{\gamma}-\rangle|\zeta_{\delta}-\rangle.\notag
\end{gather}
To shorten the writing of these wave functions, we use below the following notations:
 \begin{equation}
  \label{EQ02}
|\zeta_{\alpha} \sigma_\zeta\rangle, |\zeta_{\beta} \sigma_\zeta\rangle,|\zeta_{\gamma} \sigma_\zeta\rangle,|\zeta_{\delta} \sigma_\zeta\rangle,
 \end{equation}
 where $\sigma_\zeta = \pm 1/2$ are the spin quantum numbers in the excited state.

\subsection{Resonant energy transfer in the system of exchange-coupled spins}
 
 We define the superexchange interaction between the neighboring copper ions through intermediate boron-oxygen tetrahedrons~\cite{Ripoll1971} in the form $H_{ex} = J_{\alpha,\gamma}(\textbf{S}_\alpha\textbf{S}_\gamma)$, where $J_{\epsilon,\epsilon} = J_{\alpha,\gamma} = J_{\alpha,\delta} = J_{{\beta},\gamma} = J_{{\beta},\delta} = 3.85$~meV \cite{Boehm2002,Boehm2003}. Since the wave functions of the ground and excited states are orthogonal, then according to the Goodenough~–~Kanamori~–~Anderson rules it is logical to expect that the exchange interaction between the ground  $|\epsilon\rangle = |x^2 - y^2\rangle$ state of a single Cu$^{2+}$ ion and the excited state $|\zeta\rangle = |xy\rangle$ state of other neighboring Cu$^{2+}$ ions has ferromagnetic character, that is $J_{\zeta,\epsilon} =J_{\alpha^*,\gamma} = J_{\alpha^*,\delta} = J_{\beta^*,\delta} = J_{\beta^*,\gamma} \leq 0$.  Here the index * marks the excited state.
 
\begin{figure}[t!]
\begin{center}
\includegraphics[width=8.6cm]{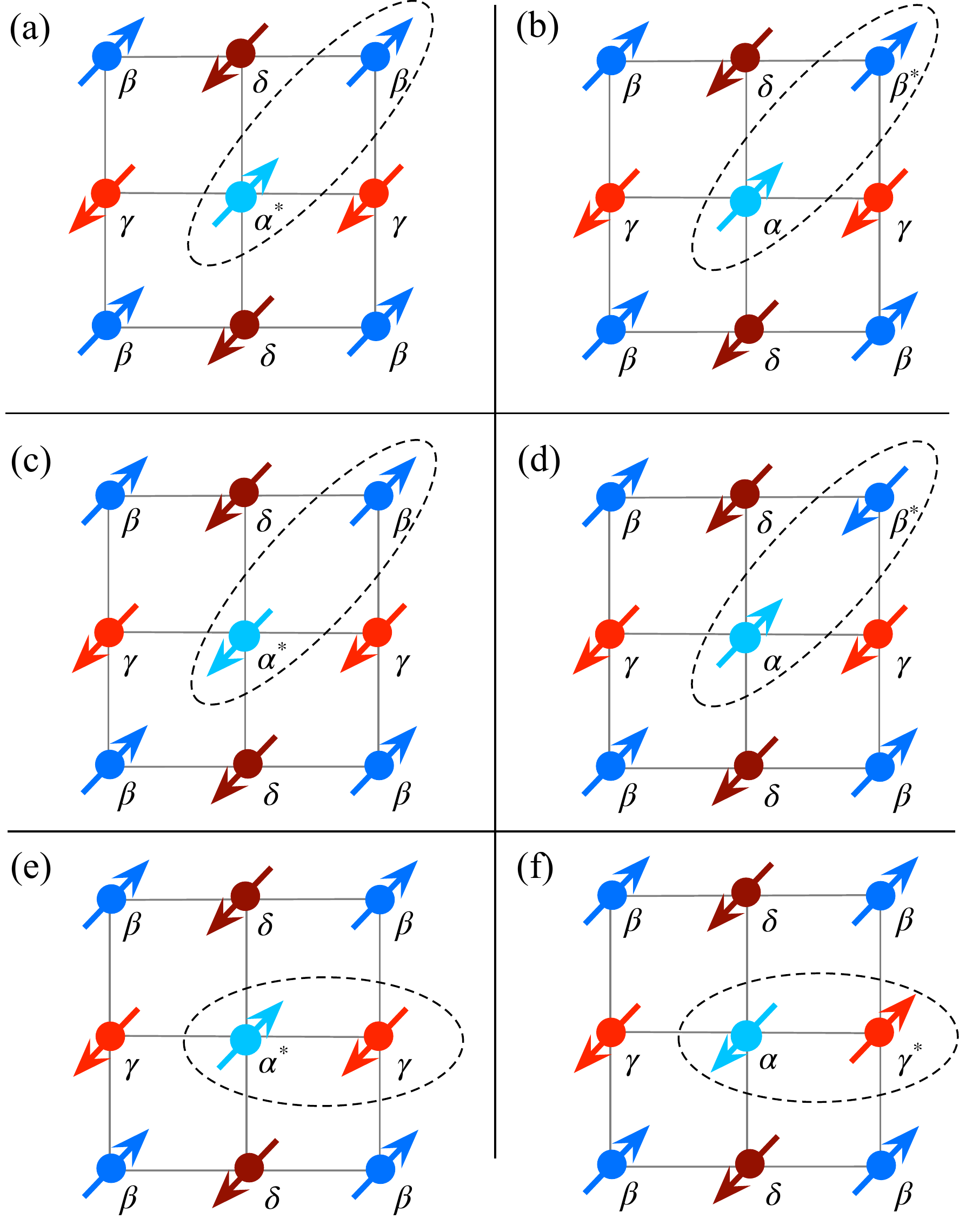}
\caption{\label{fig5} Spin structures of CuB$_2$O$_4$ under optical excitation within a single antiferromagnetic sublattice and the opposite sublattice. 
The letters $\alpha$, $\beta$, $\gamma$, and $\delta$ denote the Cu$^{2+}$ ions at the four 4$b$ positions within the single unit cell; the symbol * refers to the excited state. The pairs highlighted by dotted ovals in the (a) and (b) panels have the same excitation energy even when the exchange interaction with the neighboring ions from the opposite sublattice is taken into account. 
Thus, the resonant transfer of excitation is possible between the $\alpha$ and $\beta$ Cu$^{2+}$ ions within the same antiferromagnetic sublattice. 
Panels (c) and (d) show the spin structures under optical excitation accompanied by the spin flip. 
In this case, the resonant transfer of excitation accompanied by a spin flip is also possible. 
Panels (e) and (f) demonstrate the optical excitation of the $\alpha^*$ and $\gamma^*$ Cu$^{2+}$ ions from opposite antiferromagnetic sublattices when the excitation energy transfer is forbidden.}
\end{center}
\end{figure}

Various cases of the influence of spin ordering on the transfer of excitation energy between the Cu$^{2+}$ ions are explained in Fig.~\ref{fig5}. A pair of Cu$^{2+}$ ions at the $\alpha$ and $\beta$ positions within the same antiferromagnetic sublattice is considered in Figs.~\ref{fig5}(a) and \ref{fig5}(b). Panel (a) shows a case when under optical excitation the $\alpha$ ion moves into the $\alpha^*$ excited state  $|\zeta\rangle = |xy\rangle$ with the conservation of the spin direction, while the $\beta$ ion stays in the ground $|\epsilon\rangle~=~|x^2~-~y^2\rangle$ state. In panel (b) the case is inversed -- the $\alpha$ ion stays in the ground state, whereas the $\beta^*$ ion moves into the excited state. The optical excitation energies of the discussed pair, taking into account also the nearest Cu$^{2+}$ ions, are the same in the both cases shown in Figs.~\ref{fig5}(a) and \ref{fig5}(b), i.e. $E_{ex}(a)~=~E_{ex}(b)$:
 \begin{multline}
  \label{EQ03}
  E_{ex}(a) =-2\frac{1}{4}J_{\alpha^*,\gamma} -2\frac{1}{4}J_{\alpha^*,\delta}
  -2\frac{1}{4}J_{\beta,\gamma} -2\frac{1}{4}J_{\beta,\delta} \\= -J_{\epsilon,\epsilon}-J_{\zeta,\epsilon}.
 \end{multline} 

Thus, the equality condition for the excitation energies of Cu$^{2+}$ ions at the $\alpha$ and $\beta$ positions is not violated. Similarly, one can verify that the above condition remains valid also under optical excitation accompanied by a spin flip  because in this case the pair energy is equal to:
 \begin{equation} 
   \label{EQ04}
  E_{ex}(c)  = E_{ex}(d) = J_{\zeta,\epsilon}-J_{\epsilon,\epsilon}.
 \end{equation}
Here, $E_{ex}(c)$ and $E_{ex}(d)$ are the optical excitation energies for the configurations shown in Figs.~\ref{fig5}(c) and \ref{fig5}(d), respectively.
 
Let us now discuss the processes of optical excitation when the exchange interaction takes place between ions from opposite antiferromagnetic sublattices. Figures~\ref{fig5}(e) and \ref{fig5}(f) demonstrate an example of a pair of ions at the $\alpha$ and $\gamma$ positions. Similar to the previous case, $\alpha$, $\gamma$ and $\alpha^*$, $\gamma^*$ mark the Cu$^{2+}$ ions in the ground and excited state, respectively. The exchange interaction energy between spins for the two configurations shown in Figs.~\ref{fig5}(e) and \ref{fig5}(f) can be written as:
    \begin{equation} 
   \label{EQ05}
  E_{ex}(e) = -\frac{3}{4}J_{\gamma,\beta} - \frac{3}{4}J_{\alpha^*,\delta} - \frac{1}{4}J_{\alpha^*,\gamma} ,
 \end{equation} 
   \begin{equation} 
   \label{EQ06}
  E_{ex}(f) = \frac{3}{4}J_{\gamma,\beta} + \frac{3}{4}J_{\alpha^*,\delta} - \frac{1}{4}J_{\alpha^*,\gamma},
 \end{equation}
where $E_{ex}(e)$ and $E_{ex}(f)$ are the optical excitation energies for the two configurations shown in Figs.~\ref{fig5}(e) and \ref{fig5}(f), respectively.  One can see, that the energies in these two cases are different and, therefore, the condition for resonant transfer of excitation is violated. Thus, we come to the conclusion that substantial resonant excitation transfer occurs only within each of the two opposite antiferromagnetic sublattices.  It should be noted that this important conclusion has not been established in previous publications. Exchange interaction causes the splitting of ground and excited states and optical excitation energies become changed. 

\subsection{Davydov splitting of Frenkel excitons}

Next, we  move on to the discussion on the interaction of the excited state of Cu$^{2+}$ ion with the surrounding Cu$^{2+}$ ions in terms of the molecular field approximation. For the Cu$^{2+}$($\alpha^*$) ion, the molecular field operator has the form $\hat{H}_\text{mol} = 2\mu_\text{B}\mathbf{S}_{\alpha^*}\mathbf{M}_{\alpha}$ in which the exchange field $\mathbf{M}_{\alpha}$ is considered as a fitting parameter.
It can be seen from Figs.~\ref{fig5}(a) or \ref{fig5}(b) that $H_\text{mol}(\alpha^*) = H_\text{mol}(\beta^*) = -J_{\zeta,\epsilon}$, but for example in Fig.~\ref{fig5}(f), $H_\text{mol}(\gamma^*) = J_{\zeta,\epsilon}$,  i.e. the exchange interaction of the excited state with surrounding spins for different positions in the unit cell is different and therefore it has to be taken into account.  

Let us further consider the mechanisms of the excitation energy transfer from one Cu$^{2+}$ ion to another using as an example the $\alpha$ and $\beta$ pair of ions. 
The energies of the states with the wave functions $\alpha_{\zeta\sigma}^{+}\beta_{\epsilon\sigma}^{+}|0\rangle$ and $\alpha_{\epsilon\sigma}^{+}\beta_{\zeta\sigma}^{+}|0\rangle$ are equal. 
The Frenkel-Davydov integral of the excitation transfer is determined by an equation:
 \begin{multline}
\label{EQ07}
t_\text{F}^{(1)} = \langle \epsilon_{\alpha}(\mathbf{r}_1)\zeta_{\beta}(\mathbf{r}_2)|V(\mathbf{r}_1 - \mathbf{R}_{\alpha,\beta} - \mathbf{r}_2)|\zeta_{\alpha}(\mathbf{r}_1)\epsilon_{\beta}(\mathbf{r}_2) \rangle\\ 
-\langle\epsilon_{\alpha}(\mathbf{r}_1)\zeta_{\beta}(\mathbf{r}_2)|V(\mathbf{r}_1 - \mathbf{R}_{\alpha,\beta} - \mathbf{r}_2)|\zeta_{\alpha}(\mathbf{r}_2)\epsilon_{\beta}(\mathbf{r}_1) \rangle,
\end{multline}
where $V(\mathbf{r}_1 - \mathbf{R}_{\alpha,\beta} - \mathbf{r}_2)$ is the Coulomb interaction of electrons,  $\mathbf{R}_{\alpha,\beta}$ is the radius-vector between the $\alpha$ and $\beta$ positions, and $\mathbf{r}_1$ and $\mathbf{r}_2$ are counted from these two positions, respectively.  The first term in this equation is the Coulomb two-center integral, and the second one is the exchange integral.

\begin{figure}[t!]
\begin{center}
\includegraphics[width=8.6cm]{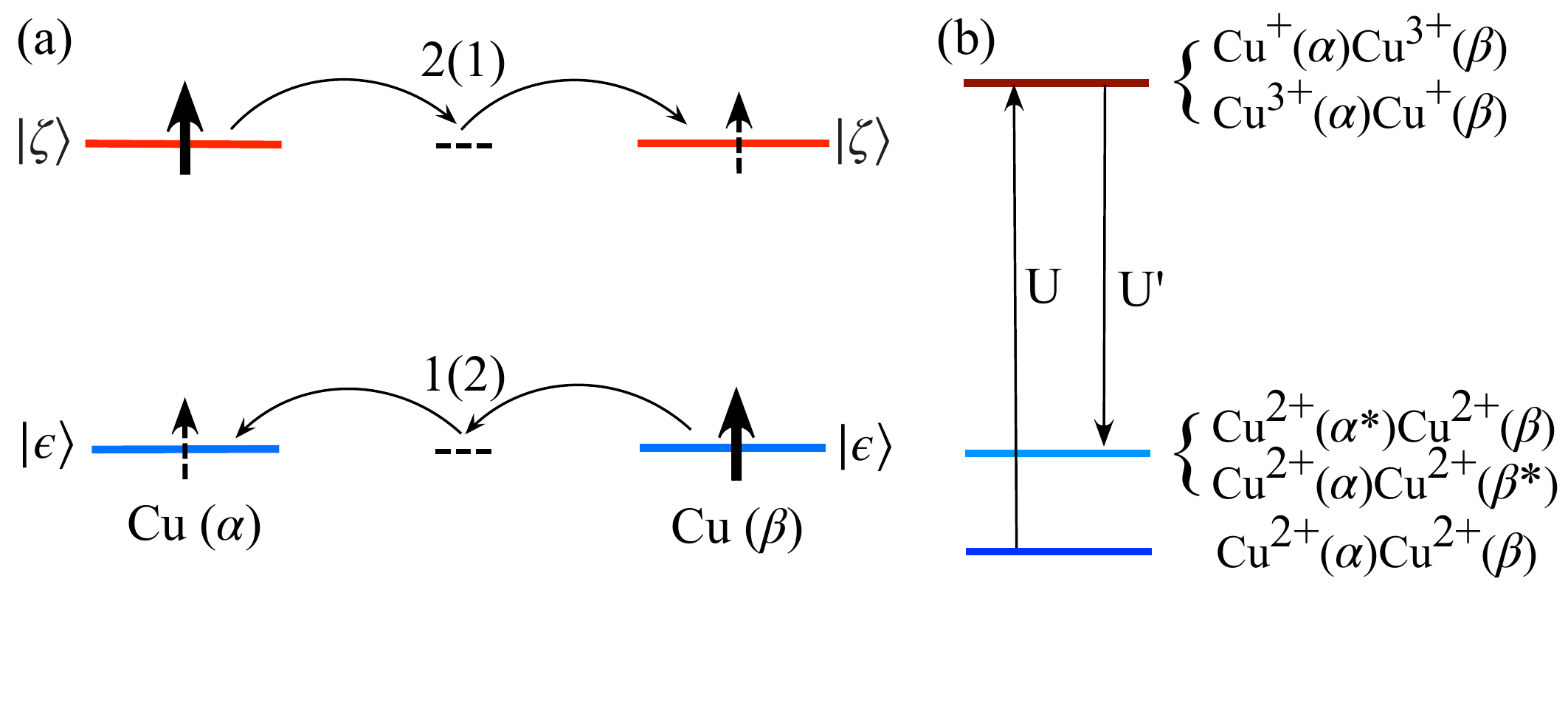}
\caption{\label{fig6} 
(a) Illustration of the excitation transfer mechanism caused by an electron hopping from one Cu$^{2+}$ ion to another through the intermediate tetrahedral BO$_4$ groups~\cite{Ripoll1971} shown schematically by the dashed lines. Thick and dashed vertical arrows show the initial and final positions for spin hopping. (b) Energy level diagram for deriving the operator of excitation transfer between the [Cu$^{2+}(\alpha)$ Cu$^{2+}(\beta)$] and [Cu$^{2+}(\alpha^*)$ Cu$^{2+}(\beta)$] or [Cu$^{2+}(\alpha)$ Cu$^{2+}(\beta^*)$] states through the  [Cu$^{+}(\alpha)$ Cu$^{3+}(\beta)$] or [Cu$^{3+}(\alpha)$ Cu$^{+}(\beta)$] states.}
\end{center}
\end{figure}

Figure~\ref{fig6}(a) shows a diagram of the excitation transfer from one Cu$^{2+}$ ion to another. The numbers 1(2) and 2(1) denote the cascade hopping of electrons between the order of Cu$^{2+}$ ions through the intermediate tetrahedral BO$_4$ groups~\cite{Ripoll1971}. The matrix elements describing electron hopping between the ground and excited states are:
\begin{eqnarray}
\label{EQ08}
t_{\epsilon\epsilon} = \langle \epsilon(\mathbf{r}-\mathbf{R}_\alpha)|\frac{p^2}{2m}+V|\epsilon(\mathbf{r}-\mathbf{R}_\beta) \rangle,\\ 
\label{EQ015}
t_{\zeta\zeta} = \langle \zeta(\mathbf{r}-\mathbf{R}_\alpha)|\frac{p^2}{2m}+V|\zeta(\mathbf{r}-\mathbf{R}_\beta) \rangle.
\end{eqnarray}
Here, $\mathbf{r}$ is a variable for integration over space, $\mathbf{R}_\alpha$ and $\mathbf{R}_\beta$ are the radius vectors to the Cu$^{2+}$ ion in $\alpha$ and $\beta$ positions, respectively, $p$ is the momentum of electron and $m$ is an electron mass. These equations show that the matrix elements include the kinetic energy operator and that the ground and excited states are independent of spin variables. In this regard, by analogy with the Anderson theory of superexchange interaction, this mechanism of excitation transfer can be classified as kinetic~\cite{Anderson1959}. The energy of electron transfer from one Cu$^{2+}$ ion to another is $E = 5.9$~eV ~\cite{Anderson1959} and it is larger than optical excitation energy $E = 1.40575$~eV. Therefore, in the second order of perturbation theory we can obtain the following equation for the transfer integral
\begin{equation}
\label{EQ09}
t_F^{(2)} = \left[\frac{1}{U} + \frac{1}{U'} \right]t_{\epsilon\epsilon}t_{\zeta\zeta}.
\end{equation}
Here $U$ is the transfer energy of an electron from the Cu$^{2+}(\alpha)$ state to the [Cu$^{+}(\alpha)$ Cu$^{3+}(\beta)$] state and $U'$ is the transfer energy of an electron from the [Cu$^{+}(\alpha)$ Cu$^{3+}(\beta)$] state to Cu$^{2+}(\alpha^*)$. The electron transfer between the Cu$^{2+}(\beta)$ and Cu$^{2+}(\beta^*)$ states through the [Cu$^{3+}(\alpha)$ Cu$^{+}(\beta)$] state can be described by the $U$ and $U'$ energies. An explanatory diagram of the virtual excitations is shown in Fig.~\ref{fig6}(b).  When deriving Eq.~(\ref{EQ09}), it is assumed that $U - U'\ll U$.

Taking into account the spatial distribution of the phases of the involved wave functions of the ground $|\epsilon_a\rangle$ and excited $|\zeta_a\rangle$ states, one can make sure that the product $t_{\epsilon\epsilon}t_{\zeta\zeta}<0$. The total transfer integral $t_F = t_F^{(1)} + t_F^{(2)}$ is considered as an adjustable parameter which is determined from experimental data. 
The $t_F^{(1)}$ value calculated using the Hartree-Fock wave functions of the Cu$^{2+}$(3$d^9$) ion is about $-6\times 10^{-6}$ meV.   
Since the absolute value of $t_\text{F}=0.06$~meV determined from experimental data turned out to be much larger than the calculated value we conclude that $t_F^{(2)}<0$ plays the dominant role and thus the total $t_F<0$. The Davydov splitting defined from experiment is equal to $\Delta_D = 2t_\text{F}$. 

\subsection{Frenkel excitons in an applied magnetic field}

Interaction of the magnetic Cu$^{2+}$ ions ($S$~=~1/2) with an applied magnetic field results in the Zeeman splitting of the ground $|\epsilon\rangle$ and excited $|\zeta\rangle$ states (we remind that  $S$~=~1/2 in both these states) and the relevant splitting is described in the usual way by operators of the form:
\begin{eqnarray}
\label{EQ10}
\hat{H}_\text{Z}^{\epsilon(\zeta)} = \mu_\text{B}g^{\epsilon(\zeta)}_{l}S^{\epsilon(\zeta)}_{l}B_{l}.
\end{eqnarray}
 Here, $\mu_\text{B}$ is the Bohr magneton, the indices $l = (a, b, c)$ define the crystallographic axes, $g^{\epsilon(\zeta)}_{l}$ is the effective $g$-factor of the Cu$^{2+}$ ion that depends on the magnetic susceptibility tensor $\chi_{ll}$, and $B_{l}$ is the applied magnetic field.

The energy matrix in the basis of the $\psi_i$ ($i = 1-8$) states from Eqs.~(\ref{EQ01}) has the following nonzero elements for the $\textbf{B} \parallel a$ geometry:
\begin{gather}
\label{EQ11}
 H(1,1) = - \Delta^{\zeta}/2 + \mu_\text{B}(g_a^\zeta - g_a^\epsilon)H_a/2,\\
 H(2,2) = -\Delta^{\zeta}/2  - \mu_\text{B}(g_a^\zeta + g_a^\epsilon)H_a/2,\\
 H(3,3) = H(1,1),\\
 H(4,4) = H(2,2),\\
 H(5,5) = \Delta^{\zeta}/2  + \mu_\text{B}(g_a^\zeta + g_a^\epsilon)H_a/2,\\
 H(6,6) = \Delta^{\zeta}/2 - \mu_\text{B}(g_a^\zeta - g_a^\epsilon)H_a/2,\\
 H(7,7) = H(5,5),\\
 H(8,8) = H(6,6),\\
 H(1,3) = H(2,4) = H(5,7) = H(6,8) =t_{\text{F}}.
\end{gather}
The spin quantization axis is chosen along the applied magnetic field $H_a$, $B_{a} = (1+4\pi\chi_{aa})H_{a}$ is the magnetic field induction component, and $\chi_{aa}$ is the magnetic susceptibility tensor component. For describing the experimental data, the values $g_\text{a}^\zeta$ and $g_\text{a}^\epsilon$, which include the ($1+4\pi\chi_{aa}$) term, are considered as fitting parameters. It is assumed that they do not vary in the entire range of magnetic fields. In the case $\textbf{H}\parallel c$, the $a$ index should be replaced by the $c$ index in the given matrix elements. The diagonalization of the $H(i,j)$ matrix makes it possible to find the relative position of the exciton energy levels while the average energy value remains the same when summing over all states as in the absence of the magnetic field.
 
 For obtaining the energies of the optical transitions, the difference $E(\zeta)-E(\epsilon)+\Delta_l^{\epsilon}$ should be added to all diagonal elements of the $H(i,j)$ matrix. Here, the quantity $\Delta_l^{\epsilon} = \mu_B (1/2 g_l^\epsilon + \delta g_l) B_l$ takes into account the energy decrease of the lowest Zeeman component of the copper ion in the ground state from which the optical transitions occur. The parameter $\delta g_l$ is an additional factor introduced for better fitting results. Origin of $\delta g_l$ is not clear. The energy value $E(\zeta)-E(\epsilon)=1.40556$~eV is due to the action of the crystal and exchange fields, as well as the spin-orbit interaction, on the Cu$^{2+}$ 4$b$ ion in zero magnetic field. 

As shown before, the developed microscopic model made it possible to give a substantiated and consistent description of the experimental results. The details of the modeling and the parameters obtained from the simulation are given in the next Sec.~\ref{modeling}.

\section{Modeling of experimental results}
\label{modeling}

In this Section, we use the developed theory for modelling our experimental results on the behavior of Frenkel excitons in magnetic field. The modelling allows us to evaluate the exciton parameters which control the energy spectra and their modifications with increasing magnetic field. The best fits to the experimental data at $T=1.6$~K are shown by the lines in Figs.~\ref{fig2}(c) and \ref{fig2}(f). The fit parameters are given in the Figure caption and also in Table~\ref{tab:t3}. Only five parameters are used for the fitting, however they are related to different properties of the exciton structure and, therefore, each of them can be evaluated with high accuracy. The splitting  $\Delta^\zeta=0.50$~meV between the A and B broad lines in zero magnetic field is due to the exchange interaction of the excited state of the Cu$^{2+}$ ion with the surrounding Cu$^{2+}$ ions from both the 4$b$ and 8$d$ subsystems. The Davydov splitting of $\Delta_{\rm D}=0.12$~meV within doublets is independent of the magnetic field and provides the transfer integral value of $t_\text{F}=\Delta_{\rm D}/2=0.06$~meV. The slopes of the four doublets in magnetic field give the $g$-factors of the ground ($g^\epsilon$) and excited ($g^\zeta$) states and also the additional spectral shift of all spectral lines due to the splitting of the ground state $\Delta^{\epsilon}_l$. The parameter $\Delta^{\epsilon}_l$ is responsible for the shift of the center of gravity of the two A and B doublet sets.   

We perform fits for all sets of experimental data for the two field configurations and four temperatures which are presented above. For evaluation it is worthwhile to give the equations for the slopes of the exciton lines in magnetic field which are given by the $g$-factors in the ground and excited state and the additional correction factor $\delta g_l$. One can define the $g$-factors from the slopes of the spectral lines given in Table~\ref{tab:t1}:
\begin{eqnarray}
\label{EQ12}
\beta(A_{1,2}) = \delta g_l -
g_{l}^\zeta/2,\\
\label{EQ13}
\beta(A_{3,4}) = \delta g_l + g_{l}^\zeta/2,\\
\label{EQ14}
\beta(B_{5,6}) = \delta g_l - g_{l}^\zeta/2 + g_{l}^\epsilon,\\
\label{EQ15}
\beta(B_{7,8}) = \delta g_l + g_{l}^\zeta/2 + g_{l}^\epsilon.
\end{eqnarray}

The evaluated experimental values $\Delta^\zeta$, $t_\text{F}$ and the calculated parameters $g^\epsilon$, $g^\zeta$ and $\delta g$ are collected in Table~\ref{tab:t3}. We conclude from the presented values that a temperature increase up to $T=10$~K does not change the splitting $\Delta^\zeta$ due to the exchange interaction with surrounding copper ions.  However, the Davydov splitting $\Delta_D = 2t_F$ is increased by a factor of 1.5 and 1.3 for the $\textbf{B} \parallel c$ and  $\textbf{B} \parallel a$ geometry, respectively. Along with this, the $g$-factors of the ground and excited states are increased  approximately by value of 0.6. The additional $\delta g$ factor changes with temperature. A further increase of the temperature complicates the determination of the Davydov splitting, since the line widths are increased considerably and only two broad lines remain out of the eight narrow exciton lines. To determine the $g$-factor in the excited state, knowledge of the slope of at least three exciton line pairs is required allowing to solve Eqs.~(\ref{EQ12}-\ref{EQ15}) with the three unknown variables $g^\epsilon$, $g^\zeta$, and $\delta g$. From $T$~=~10~K upwards, only $g^\epsilon$ can be calculated. $g^\epsilon$ changes between 1.86 and 1.58 for $\textbf{B} \parallel c$ and 2.88 and 2.56 for $\textbf{B} \parallel a$. The parameter $\Delta^\zeta$ stays constant but decreases to $\Delta^{\zeta} = 0.4$~meV at $T$~=~16~K.

\begin{table}[t!]
\caption{Experimental parameters $\Delta^\zeta$, $t_\text{F}$  and calculated values of $g^\epsilon$, $g^\zeta$ and $\delta g$ at different temperatures in $\textbf{B}\parallel c$ and $\textbf{B}\parallel a$.}
\label{tab:t3}
\begin{center}
\begin{tabular*}{0.46\textwidth}{@{\extracolsep{\fill}} |>{\centering\arraybackslash} m{0.06\textwidth}|>{\centering\arraybackslash} m{0.08\textwidth}|>{\centering\arraybackslash} m{0.08\textwidth}|>{\centering\arraybackslash} m{0.06\textwidth}|>{\centering\arraybackslash} m{0.06\textwidth}|>{\centering\arraybackslash} m{0.06\textwidth}|}
\hline
$T$ (K) & $\Delta^\zeta$ (meV) & $t_\text{F}$ (meV) & $g_l^\epsilon$ & $g_l^\zeta$ & $\delta g_l$\\
\hline
\multicolumn{6}{|c|}{$\textbf{B} \parallel c$}\\
\hline
1.6 & 0.50 & 0.06 & 1.25 & 1.81 & 0.30\\
\hline
6.0 & 0.50 & 0.09 & 1.86 & 2.02 & 0.23 \\
\hline
10.0 & 0.50 &    & 1.87 &   &  \\
\hline
16.0 & 0.40 &    & 1.58 &   &  \\
\hline
\multicolumn{6}{|c|}{$\textbf{B} \parallel a$}\\
\hline
1.6 & 0.50 & 0.06  & 2.06 & 1.93 & -0.04\\
\hline
6.0 & 0.50 & 0.08 & 2.73 & 2.09 & -0.28\\
\hline
10.0 & 0.50 &    & 2.88 &   &  \\
\hline
16.0 & 0.40 &   & 2.56 &   &  \\
\hline
\end{tabular*}
\end{center}
\end{table}

In addition to the parameters of the Frenkel excitons extracted from the modeling, it is also necessary to comment on the different intensities of the spectral lines observed in the experiment. The intensity of the spectral lines is determined by the ED or MD character of the exciton transition. Therefore we calculated the wave functions belonging to $q = 0$ exciton states with energies at $B = 8$~T and they are collected in Table~\ref{tab:t2}. We note that the wave functions have the same form for $B > 1.6$~T when CuB$_2$O$_4$ is in the commensurate phase \cite{Pankrats2018}. The wave functions are given for the exciton bands in the crystal structure with an inversion center. In such a case, a set of purely symmetric (upper energy component of doublets) and purely antisymmetric (lower component) exciton states should be observed. Considering that the lowest (A$_{1}$) state is symmetric, we come to the conclusion that optical transitions to the upper doublet components would have the ED character, while the lower ones would have the MD character. Note that if the sign of the exchange field in the excited state is changed to the opposite, then the symmetric and antisymmetric states in the doublets swap places with each other. However, we should note that the actual crystal structure of CuB$_2$O$_4$ is noncentrosymmetric (point group $-42m$) and, therefore, the absorption lines observed in experiments might have mixed ED-MD character.

\begin{table}[b!]
\caption{Wave functions $\psi_i$ of Frenkel exciton states X with the wave vector $q=0$ excited by photons with $E_{X}$ energy at $B = 8$~T for the $\textbf{B} \parallel a$ geometry.}
\label{tab:t2}
\begin{center}
\begin{tabular*}{0.37\textwidth}{@{\extracolsep{\fill}} |>{\centering\arraybackslash} m{0.06\textwidth}|>{\centering\arraybackslash} m{0.09\textwidth}|>{\centering\arraybackslash} m{0.18\textwidth}|}
\hline
X & $E_{\text{X}}$~(eV) & Wave functions $\psi_i$ \\
\hline
A$_{1}$ & 1.40510 & $(|\zeta_{\alpha} - \rangle + |\zeta_{\beta}-\rangle)/\sqrt{2}$ \\
\hline
A$_{2}$ & 1.40520 & $(|\zeta_{\alpha} - \rangle - |\zeta_{\beta}-\rangle)/\sqrt{2}$ \\
\hline
A$_{3}$ & 1.40600 & $(|\zeta_{\alpha} + \rangle + |\zeta_{\beta}+\rangle)/\sqrt{2}$ \\
\hline
A$_{4}$ & 1.40610 & $(|\zeta_{\alpha} + \rangle - |\zeta_{\beta}+\rangle)/\sqrt{2}$ \\
\hline
B$_{5}$ & 1.40656 & $(|\zeta_{\gamma}-\rangle + |\zeta_{\delta}-\rangle)/\sqrt{2}$ \\
\hline
B$_{6}$ & 1.40666 & $(|\zeta_{\gamma}-\rangle - |\zeta_{\delta}-\rangle)/\sqrt{2}$ \\
\hline
B$_{7}$ & 1.40745 & $(|\zeta_{\gamma}+\rangle + |\zeta_{\delta}+\rangle)/\sqrt{2}$ \\
\hline
B$_{8}$ & 1.40755 & $(|\zeta_{\gamma}+\rangle - |\zeta_{\delta}+\rangle)/\sqrt{2}$ \\
\hline
\end{tabular*}
\end{center}
\end{table}

\section{Conclusions}
\label{Conclusions}

We carried out a detailed experimental study of polarized magneto-absorption on the 4$b$ subsystem of the Cu$^{2+}$ ions in the CuB$_2$O$_4$ antiferromagnet. 
For fulfilling this task, we used optical spectroscopy with high spectral resolution in strong magnetic fields up to 9.5~T in the temperature range from $T=1.6$~K up to the  antiferromagnetic-paramagnetic phase transition at $T_N=20$~K. 
The study was performed in the range of the lowest-in-energy electronic transition between the orbitally nondegenerate ground $|\epsilon\rangle = |x^2-y^2\rangle$ and excited $|\zeta\rangle = |xy\rangle$ states of the Cu$^{2+}$ ion in the spectral range of 1.4055--1.4065~eV.
Though this transition was previously studied in several publications cited in the introductory Sec.~\ref{Introduction}, our approach with the use of high spectral resolution and magnetic field allowed us to get completely unexpected results. At the lowest temperature $T=1.6$~K, only a doublet of broad lines with the partially resolved splitting of 0.50 meV was observed at zero magnetic field. This splitting is unexpected because the involved electronic transition takes place between orbitally nondegenerate states. It was assumed earlier in Ref.~\cite{Boldyrev2015} that the doublet structure is due to the Davydov splitting originating from the presence of the two Cu$^{2+}$ 4$b$ ions in the primitive unit cell of CuB$_2$O$_4$.  
However, above some critical magnetic field $B_0 = 1.4$~T we discovered a well resolved splitting of the zero field doublet into a fan of eight narrow lines and these observations required a serious revision of the assumption put forward in Ref.~\cite{Boldyrev2015}.
Thus, our intriguing observations put on the agenda the question of the origin of the observed structure of Frenkel excitons which are subject to Zeeman and Davydov splitting. Moreover, our results have shown that the Davydov splitting of Frenkel excitons in CuB$_2$O$_4$ is radically different from previous studies of such splitting in the Cr$^{3+}$ based antiferromagnets in which only a part of the expected whole set of Frenkel excitons was observed as discussed in Sec.~\ref{Introduction}.

The theoretical model developed in the present paper is based on a consistent analysis of the crystallographic and magnetic symmetry of the commensurate antiferromagnetic structure of the 4$b$ spin Cu$^{2+}$ subsystem and exchange interactions within the 4$b$ subsystem and its interactions with the 8$d$ subsystem with oppositely oriented spins. As a result, the theoretical model allowed a convincing confirmation of the experimentally observed structure of Zeeman and Davydov splitting of Frenkel excitons.
Within the framework of the developed theory, reasonable values of the related parameters were obtained such as the exchange splitting $\Delta^\zeta$~=~0.5 meV of the excited state, the Davydov splitting of $\Delta_D~=~2t_F$~=~0.12~meV, where $t_F$ is the integral of the excitation transfer between the Cu$^{2+}$ ion within the unit cell, as well as the $g$ factors of the ground and excited states of the Cu$^{2+}$ 4$b$ ions, including the $g$-factor anisotropy. The experimental and theoretical approach applied for solving complex exciton spectra in CuB$_2$O$_4$ can be also used in the studies of other antiferromagnets in which  Frenkel excitons give decisive contributions to their optical spectra.

\textbf{Acknowledgments} 
This work was supported by the Deutsche Forschungsgemeinschaft via the International Collaborative Research Centre TRR 160 (Projects B2 and C8). The contribution of R.V.P. to this work was supported by Russian Foundation for Basic Research, Project No. 19-52-12063. The theoretical work of M.V.E. and A.R.N. is supported by Russian Science Foundation, Project No. 19-12-00244.


\begin{thebibliography}{45}%
\makeatletter
\providecommand \@ifxundefined [1]{%
 \@ifx{#1\undefined}
}%
\providecommand \@ifnum [1]{%
 \ifnum #1\expandafter \@firstoftwo
 \else \expandafter \@secondoftwo
 \fi
}%
\providecommand \@ifx [1]{%
 \ifx #1\expandafter \@firstoftwo
 \else \expandafter \@secondoftwo
 \fi
}%
\providecommand \natexlab [1]{#1}%
\providecommand \enquote  [1]{``#1''}%
\providecommand \bibnamefont  [1]{#1}%
\providecommand \bibfnamefont [1]{#1}%
\providecommand \citenamefont [1]{#1}%
\providecommand \href@noop [0]{\@secondoftwo}%
\providecommand \href [0]{\begingroup \@sanitize@url \@href}%
\providecommand \@href[1]{\@@startlink{#1}\@@href}%
\providecommand \@@href[1]{\endgroup#1\@@endlink}%
\providecommand \@sanitize@url [0]{\catcode `\\12\catcode `\$12\catcode
  `\&12\catcode `\#12\catcode `\^12\catcode `\_12\catcode `\%12\relax}%
\providecommand \@@startlink[1]{}%
\providecommand \@@endlink[0]{}%
\providecommand \url  [0]{\begingroup\@sanitize@url \@url }%
\providecommand \@url [1]{\endgroup\@href {#1}{\urlprefix }}%
\providecommand \urlprefix  [0]{URL }%
\providecommand \Eprint [0]{\href }%
\providecommand \doibase [0]{https://doi.org/}%
\providecommand \selectlanguage [0]{\@gobble}%
\providecommand \bibinfo  [0]{\@secondoftwo}%
\providecommand \bibfield  [0]{\@secondoftwo}%
\providecommand \translation [1]{[#1]}%
\providecommand \BibitemOpen [0]{}%
\providecommand \bibitemStop [0]{}%
\providecommand \bibitemNoStop [0]{.\EOS\space}%
\providecommand \EOS [0]{\spacefactor3000\relax}%
\providecommand \BibitemShut  [1]{\csname bibitem#1\endcsname}%
\let\auto@bib@innerbib\@empty
\bibitem [{\citenamefont {Mendeleev}(1900)}]{Mendeleev:1900}%
  \BibitemOpen
  \bibfield  {author} {\bibinfo {author} {\bibfnamefont {D.~I.}\ \bibnamefont
  {Mendeleev}},\ }\href@noop {} {\emph {\bibinfo {title} {Foundations of
  Chemistry}}}\ (\bibinfo  {publisher} {Saint Petersburg},\ \bibinfo {year}
  {1900})\BibitemShut {NoStop}%
\bibitem [{\citenamefont {Martinez-Ripoll}\ \emph {et~al.}(1971)\citenamefont
  {Martinez-Ripoll}, \citenamefont {Martinez-Carrera},\ and\ \citenamefont
  {Garcia-Blanco}}]{Ripoll1971}%
  \BibitemOpen
  \bibfield  {author} {\bibinfo {author} {\bibfnamefont {M.}~\bibnamefont
  {Martinez-Ripoll}}, \bibinfo {author} {\bibfnamefont {S.}~\bibnamefont
  {Martinez-Carrera}},\ and\ \bibinfo {author} {\bibfnamefont {S.}~\bibnamefont
  {Garcia-Blanco}},\ }\bibfield  {title} {\bibinfo {title} {The crystal
  structure of copper metaborate {CuB}$_2${O}$_4$},\ }\href@noop {} {\bibfield
  {journal} {\bibinfo  {journal} {Acta Cryst. Sec. B}\ }\textbf {\bibinfo
  {volume} {27}},\ \bibinfo {pages} {677} (\bibinfo {year} {1971})}\BibitemShut
  {NoStop}%
\bibitem [{\citenamefont {Abdullaev}\ and\ \citenamefont
  {Mamedov}(1981)}]{Abdullaev1981}%
  \BibitemOpen
  \bibfield  {author} {\bibinfo {author} {\bibfnamefont {G.~K.}\ \bibnamefont
  {Abdullaev}}\ and\ \bibinfo {author} {\bibfnamefont {K.~S.}\ \bibnamefont
  {Mamedov}},\ }\bibfield  {title} {\bibinfo {title} {Refined crystal structure
  of copper metaborate {CuB}$_2${O}$_4$},\ }\href@noop {} {\bibfield  {journal}
  {\bibinfo  {journal} {J. Struct. Chem.}\ }\textbf {\bibinfo {volume} {22}},\
  \bibinfo {pages} {637} (\bibinfo {year} {1981})}\BibitemShut {NoStop}%
\bibitem [{\citenamefont {Petrova}\ and\ \citenamefont
  {Pankrats}(2018)}]{Pankrats2018}%
  \BibitemOpen
  \bibfield  {author} {\bibinfo {author} {\bibfnamefont {A.~E.}\ \bibnamefont
  {Petrova}}\ and\ \bibinfo {author} {\bibfnamefont {A.~I.}\ \bibnamefont
  {Pankrats}},\ }\bibfield  {title} {\bibinfo {title} {Copper metaborate
  {CuB}$_2${O}$_4$ phase diagrams based on the results of measuring the
  magnetic moment},\ }\href@noop {} {\bibfield  {journal} {\bibinfo  {journal}
  {J. Exp. Theor. Phys.}\ }\textbf {\bibinfo {volume} {126}},\ \bibinfo {pages}
  {506} (\bibinfo {year} {2018})}\BibitemShut {NoStop}%
\bibitem [{\citenamefont {Boehm}\ \emph {et~al.}(2003)\citenamefont {Boehm},
  \citenamefont {Roessli}, \citenamefont {Schefer}, \citenamefont {Wills},
  \citenamefont {Ouladdiaf}, \citenamefont {Leli{\`e}vre-Berna}, \citenamefont
  {Staub},\ and\ \citenamefont {Petrakovskii}}]{Boehm2003}%
  \BibitemOpen
  \bibfield  {author} {\bibinfo {author} {\bibfnamefont {M.}~\bibnamefont
  {Boehm}}, \bibinfo {author} {\bibfnamefont {B.}~\bibnamefont {Roessli}},
  \bibinfo {author} {\bibfnamefont {J.}~\bibnamefont {Schefer}}, \bibinfo
  {author} {\bibfnamefont {A.~S.}\ \bibnamefont {Wills}}, \bibinfo {author}
  {\bibfnamefont {B.}~\bibnamefont {Ouladdiaf}}, \bibinfo {author}
  {\bibfnamefont {E.}~\bibnamefont {Leli{\`e}vre-Berna}}, \bibinfo {author}
  {\bibfnamefont {U.}~\bibnamefont {Staub}},\ and\ \bibinfo {author}
  {\bibfnamefont {G.~A.}\ \bibnamefont {Petrakovskii}},\ }\bibfield  {title}
  {\bibinfo {title} {Complex magnetic ground state of {CuB}$_2${O}$_4$},\
  }\href@noop {} {\bibfield  {journal} {\bibinfo  {journal} {Phys. Rev. B}\
  }\textbf {\bibinfo {volume} {68}},\ \bibinfo {pages} {024405} (\bibinfo
  {year} {2003})}\BibitemShut {NoStop}%
\bibitem [{\citenamefont {Kawamata}\ \emph {et~al.}(2019)\citenamefont
  {Kawamata}, \citenamefont {Sugawara}, \citenamefont {Haider},\ and\
  \citenamefont {Adachi}}]{Kawamata2019}%
  \BibitemOpen
  \bibfield  {author} {\bibinfo {author} {\bibfnamefont {T.}~\bibnamefont
  {Kawamata}}, \bibinfo {author} {\bibfnamefont {N.}~\bibnamefont {Sugawara}},
  \bibinfo {author} {\bibfnamefont {S.~M.}\ \bibnamefont {Haider}},\ and\
  \bibinfo {author} {\bibfnamefont {T.}~\bibnamefont {Adachi}},\ }\bibfield
  {title} {\bibinfo {title} {Thermal conductivity and magnetic phase diagram of
  {CuB}$_2${O}$_4$},\ }\href@noop {} {\bibfield  {journal} {\bibinfo  {journal}
  {J. Phys. Soc. Jpn.}\ }\textbf {\bibinfo {volume} {88}},\ \bibinfo {pages}
  {114708} (\bibinfo {year} {2019})}\BibitemShut {NoStop}%
\bibitem [{\citenamefont {Pisarev}\ \emph {et~al.}(2004)\citenamefont
  {Pisarev}, \citenamefont {S{\"a}nger}, \citenamefont {Petrakovskii},\ and\
  \citenamefont {Fiebig}}]{Pisarev2004}%
  \BibitemOpen
  \bibfield  {author} {\bibinfo {author} {\bibfnamefont {R.~V.}\ \bibnamefont
  {Pisarev}}, \bibinfo {author} {\bibfnamefont {I.}~\bibnamefont {S{\"a}nger}},
  \bibinfo {author} {\bibfnamefont {G.~A.}\ \bibnamefont {Petrakovskii}},\ and\
  \bibinfo {author} {\bibfnamefont {M.}~\bibnamefont {Fiebig}},\ }\bibfield
  {title} {\bibinfo {title} {Magnetic-field induced second harmonic generation
  in {CuB}$_2${O}$_4$},\ }\href@noop {} {\bibfield  {journal} {\bibinfo
  {journal} {Phys. Rev. Lett.}\ }\textbf {\bibinfo {volume} {93}},\ \bibinfo
  {pages} {037204} (\bibinfo {year} {2004})}\BibitemShut {NoStop}%
\bibitem [{\citenamefont {Pisarev}\ \emph {et~al.}(2011)\citenamefont
  {Pisarev}, \citenamefont {Kalashnikova}, \citenamefont {Sch{\"o}ps},\ and\
  \citenamefont {Bezmaternykh}}]{Pisarev2011}%
  \BibitemOpen
  \bibfield  {author} {\bibinfo {author} {\bibfnamefont {R.~V.}\ \bibnamefont
  {Pisarev}}, \bibinfo {author} {\bibfnamefont {A.~M.}\ \bibnamefont
  {Kalashnikova}}, \bibinfo {author} {\bibfnamefont {O.}~\bibnamefont
  {Sch{\"o}ps}},\ and\ \bibinfo {author} {\bibfnamefont {L.~N.}\ \bibnamefont
  {Bezmaternykh}},\ }\bibfield  {title} {\bibinfo {title} {Electronic
  transitions and genuine crystal-field parameters in copper metaborate
  {CuB}$_2${O}$_4$},\ }\href@noop {} {\bibfield  {journal} {\bibinfo  {journal}
  {Phys. Rev. B}\ }\textbf {\bibinfo {volume} {84}},\ \bibinfo {pages} {075160}
  (\bibinfo {year} {2011})}\BibitemShut {NoStop}%
\bibitem [{\citenamefont {Fiebig}\ \emph {et~al.}(2003)\citenamefont {Fiebig},
  \citenamefont {S{\"a}nger},\ and\ \citenamefont {Pisarev}}]{Fiebig2003}%
  \BibitemOpen
  \bibfield  {author} {\bibinfo {author} {\bibfnamefont {M.}~\bibnamefont
  {Fiebig}}, \bibinfo {author} {\bibfnamefont {I.}~\bibnamefont {S{\"a}nger}},\
  and\ \bibinfo {author} {\bibfnamefont {R.~V.}\ \bibnamefont {Pisarev}},\
  }\bibfield  {title} {\bibinfo {title} {Magnetic phase diagram of
  {CuB}$_2${O}$_4$},\ }\href@noop {} {\bibfield  {journal} {\bibinfo  {journal}
  {J. Appl. Phys.}\ }\textbf {\bibinfo {volume} {93}},\ \bibinfo {pages} {6960}
  (\bibinfo {year} {2003})}\BibitemShut {NoStop}%
\bibitem [{\citenamefont {Saito}\ \emph {et~al.}(2008)\citenamefont {Saito},
  \citenamefont {Taniguchi},\ and\ \citenamefont {Arima}}]{Saito2008}%
  \BibitemOpen
  \bibfield  {author} {\bibinfo {author} {\bibfnamefont {M.}~\bibnamefont
  {Saito}}, \bibinfo {author} {\bibfnamefont {K.}~\bibnamefont {Taniguchi}},\
  and\ \bibinfo {author} {\bibfnamefont {T.-H.}\ \bibnamefont {Arima}},\
  }\bibfield  {title} {\bibinfo {title} {Gigantic optical magnetoelectric
  effect in {CuB}$_2${O}$_4$},\ }\href@noop {} {\bibfield  {journal} {\bibinfo
  {journal} {J. Phys. Soc. Jpn.}\ }\textbf {\bibinfo {volume} {77}},\ \bibinfo
  {pages} {013705} (\bibinfo {year} {2008})}\BibitemShut {NoStop}%
\bibitem [{\citenamefont {Toyoda}\ \emph {et~al.}(2015)\citenamefont {Toyoda},
  \citenamefont {Abe}, \citenamefont {Kimura}, \citenamefont {Matsuda},
  \citenamefont {Nomura}, \citenamefont {Ikeda}, \citenamefont {Takeyama},\
  and\ \citenamefont {Arima}}]{Toyoda2015}%
  \BibitemOpen
  \bibfield  {author} {\bibinfo {author} {\bibfnamefont {S.}~\bibnamefont
  {Toyoda}}, \bibinfo {author} {\bibfnamefont {N.}~\bibnamefont {Abe}},
  \bibinfo {author} {\bibfnamefont {S.}~\bibnamefont {Kimura}}, \bibinfo
  {author} {\bibfnamefont {Y.~H.}\ \bibnamefont {Matsuda}}, \bibinfo {author}
  {\bibfnamefont {T.}~\bibnamefont {Nomura}}, \bibinfo {author} {\bibfnamefont
  {A.}~\bibnamefont {Ikeda}}, \bibinfo {author} {\bibfnamefont
  {S.}~\bibnamefont {Takeyama}},\ and\ \bibinfo {author} {\bibfnamefont
  {T.}~\bibnamefont {Arima}},\ }\bibfield  {title} {\bibinfo {title} {One-way
  transparency of light in multiferroic {CuB}$_2${O}$_4$},\ }\href@noop {}
  {\bibfield  {journal} {\bibinfo  {journal} {Phys. Rev. Lett.}\ }\textbf
  {\bibinfo {volume} {115}},\ \bibinfo {pages} {267207} (\bibinfo {year}
  {2015})}\BibitemShut {NoStop}%
\bibitem [{\citenamefont {Toyoda}\ \emph {et~al.}(2019)\citenamefont {Toyoda},
  \citenamefont {N.~Abe},\ and\ \citenamefont {Arima}}]{Toyoda2019}%
  \BibitemOpen
  \bibfield  {author} {\bibinfo {author} {\bibfnamefont {S.}~\bibnamefont
  {Toyoda}}, \bibinfo {author} {\bibfnamefont {N.}~\bibnamefont {N.~Abe}},\
  and\ \bibinfo {author} {\bibfnamefont {T.}~\bibnamefont {Arima}},\ }\bibfield
   {title} {\bibinfo {title} {Nonreciprocal second harmonic generation in a
  magnetoelectric material},\ }\href@noop {} {\bibfield  {journal} {\bibinfo
  {journal} {Phys. Rev. Lett.}\ }\textbf {\bibinfo {volume} {123}},\ \bibinfo
  {pages} {077401} (\bibinfo {year} {2019})}\BibitemShut {NoStop}%
\bibitem [{\citenamefont {Toyoda}\ \emph {et~al.}(2016)\citenamefont {Toyoda},
  \citenamefont {Abe},\ and\ \citenamefont {Arima}}]{Toyoda2016}%
  \BibitemOpen
  \bibfield  {author} {\bibinfo {author} {\bibfnamefont {S.}~\bibnamefont
  {Toyoda}}, \bibinfo {author} {\bibfnamefont {N.}~\bibnamefont {Abe}},\ and\
  \bibinfo {author} {\bibfnamefont {T.}~\bibnamefont {Arima}},\ }\bibfield
  {title} {\bibinfo {title} {Gigantic directional asymmetry of luminescence in
  multiferroic {CuB}$_2${O}$_4$},\ }\href@noop {} {\bibfield  {journal}
  {\bibinfo  {journal} {Phys. Rev. B}\ }\textbf {\bibinfo {volume} {93}},\
  \bibinfo {pages} {201109(R)} (\bibinfo {year} {2016})}\BibitemShut {NoStop}%
\bibitem [{\citenamefont {Kudlacik}\ \emph {et~al.}(2020)\citenamefont
  {Kudlacik}, \citenamefont {Ivanov}, \citenamefont {Yakovlev}, \citenamefont
  {Sapega}, \citenamefont {Schindler}, \citenamefont {Debus}, \citenamefont
  {Bayer},\ and\ \citenamefont {Pisarev}}]{Kudlacik2020}%
  \BibitemOpen
  \bibfield  {author} {\bibinfo {author} {\bibfnamefont {D.}~\bibnamefont
  {Kudlacik}}, \bibinfo {author} {\bibfnamefont {V.~Y.}\ \bibnamefont
  {Ivanov}}, \bibinfo {author} {\bibfnamefont {D.~R.}\ \bibnamefont
  {Yakovlev}}, \bibinfo {author} {\bibfnamefont {V.~F.}\ \bibnamefont
  {Sapega}}, \bibinfo {author} {\bibfnamefont {J.~J.}~\bibnamefont {Schindler}},
  \bibinfo {author} {\bibfnamefont {J.}~\bibnamefont {Debus}}, \bibinfo
  {author} {\bibfnamefont {M.}~\bibnamefont {Bayer}},\ and\ \bibinfo {author}
  {\bibfnamefont {R.~V.}\ \bibnamefont {Pisarev}},\ }\bibfield  {title}
  {\bibinfo {title} {Exciton and exciton-magnon photoluminescence in the
  antiferromagnet {CuB}$_2${O}$_4$},\ }\href@noop {} {\bibfield  {journal}
  {\bibinfo  {journal} {Phys. Rev. B}\ }\textbf {\bibinfo {volume} {102}},\
  \bibinfo {pages} {035128} (\bibinfo {year} {2020})}\BibitemShut {NoStop}%
\bibitem [{\citenamefont {Mund}\ \emph {et~al.}(2021)\citenamefont {Mund},
  \citenamefont {Yakovlev}, \citenamefont {Poddubny}, \citenamefont {Dubrovin},
  \citenamefont {Bayer},\ and\ \citenamefont {Pisarev}}]{Mund2021}%
  \BibitemOpen
  \bibfield  {author} {\bibinfo {author} {\bibfnamefont {J.}~\bibnamefont
  {Mund}}, \bibinfo {author} {\bibfnamefont {D.~R.}\ \bibnamefont {Yakovlev}},
  \bibinfo {author} {\bibfnamefont {A.~N.}\ \bibnamefont {Poddubny}}, \bibinfo
  {author} {\bibfnamefont {R.~M.}\ \bibnamefont {Dubrovin}}, \bibinfo {author}
  {\bibfnamefont {M.}~\bibnamefont {Bayer}},\ and\ \bibinfo {author}
  {\bibfnamefont {R.~V.}\ \bibnamefont {Pisarev}},\ }\bibfield  {title}
  {\bibinfo {title} {Toroidal nonreciprocity of optical second harmonic
  generation},\ }\href@noop {} {\bibfield  {journal} {\bibinfo  {journal}
  {Phys. Rev. B}\ }\textbf {\bibinfo {volume} {103}},\ \bibinfo {pages}
  {L180410} (\bibinfo {year} {2021})}\BibitemShut {NoStop}%
\bibitem [{\citenamefont {Toyoda}\ \emph {et~al.}(2021)\citenamefont {Toyoda},
  \citenamefont {Fiebig}, \citenamefont {Arima}, \citenamefont {Tokura},\ and\
  \citenamefont {Ogawa}}]{Toyoda2021}%
  \BibitemOpen
  \bibfield  {author} {\bibinfo {author} {\bibfnamefont {S.}~\bibnamefont
  {Toyoda}}, \bibinfo {author} {\bibfnamefont {M.}~\bibnamefont {Fiebig}},
  \bibinfo {author} {\bibfnamefont {T.}~\bibnamefont {Arima}}, \bibinfo
  {author} {\bibfnamefont {Y.}~\bibnamefont {Tokura}},\ and\ \bibinfo {author}
  {\bibfnamefont {N.}~\bibnamefont {Ogawa}},\ }\bibfield  {title} {\bibinfo
  {title} {Nonreciprocal second harmonic generation in a magnetoelectric
  material},\ }\href@noop {} {\bibfield  {journal} {\bibinfo  {journal}
  {Science Advances}\ }\textbf {\bibinfo {volume} {7}},\ \bibinfo {pages}
  {eabe2793} (\bibinfo {year} {2021})}\BibitemShut {NoStop}%
\bibitem [{\citenamefont {Arima}(2008)}]{Arima2008}%
  \BibitemOpen
  \bibfield  {author} {\bibinfo {author} {\bibfnamefont {T.}~\bibnamefont
  {Arima}},\ }\bibfield  {title} {\bibinfo {title} {Magneto-electric optics in
  non-centrosymmetric ferromagnets},\ }\href@noop {} {\bibfield  {journal}
  {\bibinfo  {journal} {J. Phys.: Condens. Matter}\ }\textbf {\bibinfo {volume}
  {20}},\ \bibinfo {pages} {434211} (\bibinfo {year} {2008})}\BibitemShut
  {NoStop}%
\bibitem [{\citenamefont {Arima}\ and\ \citenamefont
  {Saito}(2009)}]{ArimaC2009}%
  \BibitemOpen
  \bibfield  {author} {\bibinfo {author} {\bibfnamefont {T.}~\bibnamefont
  {Arima}}\ and\ \bibinfo {author} {\bibfnamefont {M.}~\bibnamefont {Saito}},\
  }\bibfield  {title} {\bibinfo {title} {Comment on 'calculated chiral and
  magneto-electric dichroic signals for copper metaborate ({CuB}$_2${O}$_4$) in
  an applied magnetic field'},\ }\href@noop {} {\bibfield  {journal} {\bibinfo
  {journal} {J. Phys.: Condens. Matter}\ }\textbf {\bibinfo {volume} {21}},\
  \bibinfo {pages} {498001} (\bibinfo {year} {2009})}\BibitemShut {NoStop}%
\bibitem [{\citenamefont {Lovesey}\ and\ \citenamefont
  {Staub}(2009{\natexlab{a}})}]{Lovesey2009}%
  \BibitemOpen
  \bibfield  {author} {\bibinfo {author} {\bibfnamefont {S.~W.}\ \bibnamefont
  {Lovesey}}\ and\ \bibinfo {author} {\bibfnamefont {U.}~\bibnamefont
  {Staub}},\ }\bibfield  {title} {\bibinfo {title} {Calculated chiral and
  magneto-electric dichroic signals for copper metaborate ({CuB}$_2${O}$_4$) in
  an applied magnetic field},\ }\href@noop {} {\bibfield  {journal} {\bibinfo
  {journal} {J. Phys.: Condens. Matter}\ }\textbf {\bibinfo {volume} {21}},\
  \bibinfo {pages} {142201} (\bibinfo {year} {2009}{\natexlab{a}})}\BibitemShut
  {NoStop}%
\bibitem [{\citenamefont {Lovesey}\ and\ \citenamefont
  {Staub}(2009{\natexlab{b}})}]{LoveseyC2009}%
  \BibitemOpen
  \bibfield  {author} {\bibinfo {author} {\bibfnamefont {S.~W.}\ \bibnamefont
  {Lovesey}}\ and\ \bibinfo {author} {\bibfnamefont {U.}~\bibnamefont
  {Staub}},\ }\bibfield  {title} {\bibinfo {title} {Reply to comment on
  'calculated chiral and magneto-electric dichroic signals for copper
  metaborate ({CuB}$_2${O}$_4$) in an applied magnetic field'},\ }\href@noop {}
  {\bibfield  {journal} {\bibinfo  {journal} {J. Phys.: Condens. Matter}\
  }\textbf {\bibinfo {volume} {21}},\ \bibinfo {pages} {498002} (\bibinfo
  {year} {2009}{\natexlab{b}})}\BibitemShut {NoStop}%
\bibitem [{\citenamefont {Nikitchenko}\ and\ \citenamefont
  {Pisarev}(2021)}]{Nikitchenko}%
  \BibitemOpen
  \bibfield  {author} {\bibinfo {author} {\bibfnamefont {A.~I.}\ \bibnamefont
  {Nikitchenko}}\ and\ \bibinfo {author} {\bibfnamefont {R.~V.}\ \bibnamefont
  {Pisarev}},\ }\bibfield  {title} {\bibinfo {title} {Magnetic and
  antiferromagnetic nonreciprocity of light propagation in the magnetoelectric
  {CuB}$_2${O}$_4$},\ }\href@noop {} {\bibfield  {journal} {\bibinfo  {journal}
  {Accepted to Phys. Rev. B}\ }\textbf {\bibinfo {volume} {x}},\ \bibinfo
  {pages} {x} (\bibinfo {year} {2021})}\BibitemShut {NoStop}%
\bibitem [{\citenamefont {Boldyrev}\ \emph {et~al.}(2015)\citenamefont
  {Boldyrev}, \citenamefont {Pisarev}, \citenamefont {Bezmaternykh},\ and\
  \citenamefont {Popova}}]{Boldyrev2015}%
  \BibitemOpen
  \bibfield  {author} {\bibinfo {author} {\bibfnamefont {K.~N.}\ \bibnamefont
  {Boldyrev}}, \bibinfo {author} {\bibfnamefont {R.~V.}\ \bibnamefont
  {Pisarev}}, \bibinfo {author} {\bibfnamefont {L.~N.}\ \bibnamefont
  {Bezmaternykh}},\ and\ \bibinfo {author} {\bibfnamefont {M.~N.}\ \bibnamefont
  {Popova}},\ }\bibfield  {title} {\bibinfo {title} {Antiferromagnetic
  dichroism in a complex multisublattice magnetoelectric {CuB}$_2${O}$_4$},\
  }\href@noop {} {\bibfield  {journal} {\bibinfo  {journal} {Phys. Rev. Lett.}\
  }\textbf {\bibinfo {volume} {114}},\ \bibinfo {pages} {247210} (\bibinfo
  {year} {2015})}\BibitemShut {NoStop}%
\bibitem [{\citenamefont {Frenkel}(1931{\natexlab{a}})}]{Frenkel1}%
  \BibitemOpen
  \bibfield  {author} {\bibinfo {author} {\bibfnamefont {J.}~\bibnamefont
  {Frenkel}},\ }\bibfield  {title} {\bibinfo {title} {On the transformation of
  light into heat in solids (i)},\ }\href
  {https://doi.org/DOI:https://doi.org/10.1103/PhysRev.37.17} {\bibfield
  {journal} {\bibinfo  {journal} {Phys. Rev.}\ }\textbf {\bibinfo {volume}
  {37}},\ \bibinfo {pages} {17} (\bibinfo {year}
  {1931}{\natexlab{a}})}\BibitemShut {NoStop}%
\bibitem [{\citenamefont {Frenkel}(1931{\natexlab{b}})}]{Frenkel2}%
  \BibitemOpen
  \bibfield  {author} {\bibinfo {author} {\bibfnamefont {J.}~\bibnamefont
  {Frenkel}},\ }\bibfield  {title} {\bibinfo {title} {On the transformation of
  light into heat in solids (ii)},\ }\href
  {https://doi.org/10.1103/PhysRev.37.1276} {\bibfield  {journal} {\bibinfo
  {journal} {Phys. Rev.}\ }\textbf {\bibinfo {volume} {37}},\ \bibinfo {pages}
  {1276} (\bibinfo {year} {1931}{\natexlab{b}})}\BibitemShut {NoStop}%
\bibitem [{\citenamefont {Davydov}(1964)}]{Davydov1964}%
  \BibitemOpen
  \bibfield  {author} {\bibinfo {author} {\bibfnamefont {A.~S.}\ \bibnamefont
  {Davydov}},\ }\bibfield  {title} {\bibinfo {title} {The theory of molecular
  excitons},\ }\href@noop {} {\bibfield  {journal} {\bibinfo  {journal} {Phys.
  Usp.}\ }\textbf {\bibinfo {volume} {7}},\ \bibinfo {pages} {145} (\bibinfo
  {year} {1964})}\BibitemShut {NoStop}%
\bibitem [{\citenamefont {Davydov}(1971)}]{Davydov1971}%
  \BibitemOpen
  \bibfield  {author} {\bibinfo {author} {\bibfnamefont {A.~S.}\ \bibnamefont
  {Davydov}},\ }\href@noop {} {\emph {\bibinfo {title} {Theory of Molecular
  Excitons}}}\ (\bibinfo  {publisher} {Plenum Press, New York},\ \bibinfo
  {year} {1971})\BibitemShut {NoStop}%
\bibitem [{\citenamefont {Loudon}(1968)}]{Loudon1968}%
  \BibitemOpen
  \bibfield  {author} {\bibinfo {author} {\bibfnamefont {R.}~\bibnamefont
  {Loudon}},\ }\bibfield  {title} {\bibinfo {title} {Theory of infra-red and
  optical spectra of antiferromagnets},\ }\href@noop {} {\bibfield  {journal}
  {\bibinfo  {journal} {Adv. Phys.}\ }\textbf {\bibinfo {volume} {17}},\
  \bibinfo {pages} {243} (\bibinfo {year} {1968})}\BibitemShut {NoStop}%
\bibitem [{\citenamefont {van~der Ziel}(1967)}]{Ziel1967}%
  \BibitemOpen
  \bibfield  {author} {\bibinfo {author} {\bibfnamefont {J.~P.}\ \bibnamefont
  {van~der Ziel}},\ }\bibfield  {title} {\bibinfo {title} {Optical spectrum of
  antiferromagnetic {Cr}$_{2}${O}$_{3}$},\ }\href@noop {} {\bibfield  {journal}
  {\bibinfo  {journal} {Phys. Rev.}\ }\textbf {\bibinfo {volume} {161}},\
  \bibinfo {pages} {483} (\bibinfo {year} {1967})}\BibitemShut {NoStop}%
\bibitem [{\citenamefont {Aoyagi}\ \emph {et~al.}(1969)\citenamefont {Aoyagi},
  \citenamefont {Tsushima},\ and\ \citenamefont {Sugano}}]{Aoyagi1969}%
  \BibitemOpen
  \bibfield  {author} {\bibinfo {author} {\bibfnamefont {K.}~\bibnamefont
  {Aoyagi}}, \bibinfo {author} {\bibfnamefont {K.}~\bibnamefont {Tsushima}},\
  and\ \bibinfo {author} {\bibfnamefont {S.}~\bibnamefont {Sugano}},\
  }\bibfield  {title} {\bibinfo {title} {Direct observation of {D}avydov
  splitting in antiferromagnetic {Y}{Cr}{O}$_{3}$},\ }\href@noop {} {\bibfield
  {journal} {\bibinfo  {journal} {Solid State Commun.}\ }\textbf {\bibinfo
  {volume} {7}},\ \bibinfo {pages} {229} (\bibinfo {year} {1969})}\BibitemShut
  {NoStop}%
\bibitem [{\citenamefont {Meltzer}\ \emph {et~al.}(1968)\citenamefont
  {Meltzer}, \citenamefont {Chen}, \citenamefont {McClure},\ and\ \citenamefont
  {Lowepari}}]{Meltzer1968}%
  \BibitemOpen
  \bibfield  {author} {\bibinfo {author} {\bibfnamefont {R.~S.}\ \bibnamefont
  {Meltzer}}, \bibinfo {author} {\bibfnamefont {M.~Y.}\ \bibnamefont {Chen}},
  \bibinfo {author} {\bibfnamefont {D.~S.}\ \bibnamefont {McClure}},\ and\
  \bibinfo {author} {\bibfnamefont {M.}~\bibnamefont {Lowepari}},\ }\bibfield
  {title} {\bibinfo {title} {Exciton-magnon bound state in {MnF}$_2$ and the
  exciton dispersion in {MnF}$_2$ and {RbMnF}$_3$},\ }\href@noop {} {\bibfield
  {journal} {\bibinfo  {journal} {Phys. Rev. Lett.}\ }\textbf {\bibinfo
  {volume} {21}},\ \bibinfo {pages} {913} (\bibinfo {year} {1968})}\BibitemShut
  {NoStop}%
\bibitem [{\citenamefont {Allen}\ \emph {et~al.}(1969)\citenamefont {Allen},
  \citenamefont {Macfarlane},\ and\ \citenamefont {White}}]{Allen1969}%
  \BibitemOpen
  \bibfield  {author} {\bibinfo {author} {\bibfnamefont {J.~W.}\ \bibnamefont
  {Allen}}, \bibinfo {author} {\bibfnamefont {R.~M.}\ \bibnamefont
  {Macfarlane}},\ and\ \bibinfo {author} {\bibfnamefont {R.~L.}\ \bibnamefont
  {White}},\ }\bibfield  {title} {\bibinfo {title} {Magnetic {D}avydov
  splittings in the optical absorption spectrum of {Cr}$_{2}${O}$_{3}$},\
  }\href@noop {} {\bibfield  {journal} {\bibinfo  {journal} {Phys. Rev.}\
  }\textbf {\bibinfo {volume} {179}},\ \bibinfo {pages} {523} (\bibinfo {year}
  {1969})}\BibitemShut {NoStop}%
\bibitem [{\citenamefont {Allen}(1970)}]{ALLEN1970}%
  \BibitemOpen
  \bibfield  {author} {\bibinfo {author} {\bibfnamefont {J.}~\bibnamefont
  {Allen}},\ }\bibfield  {title} {\bibinfo {title} {Reinterpretation of 4{A}2
  → 2{E} exciton spectra in {YCrO}$_3$},\ }\href@noop {} {\bibfield
  {journal} {\bibinfo  {journal} {Solid State Communications}\ }\textbf
  {\bibinfo {volume} {8}},\ \bibinfo {pages} {53} (\bibinfo {year}
  {1970})}\BibitemShut {NoStop}%
\bibitem [{\citenamefont {Meltzer}(1970)}]{Meltzer1970}%
  \BibitemOpen
  \bibfield  {author} {\bibinfo {author} {\bibfnamefont {R.~S.}\ \bibnamefont
  {Meltzer}},\ }\bibfield  {title} {\bibinfo {title} {Davydov splitting in the
  optical absorption spectra of the {Cr}$^{+3}$ $^{2}${E} state in some
  rare-earth orthochromites},\ }\href@noop {} {\bibfield  {journal} {\bibinfo
  {journal} {Phys. Rev. B}\ }\textbf {\bibinfo {volume} {2}},\ \bibinfo {pages}
  {2398} (\bibinfo {year} {1970})}\BibitemShut {NoStop}%
\bibitem [{\citenamefont {Sugano}\ \emph {et~al.}(1971)\citenamefont {Sugano},
  \citenamefont {Aoyagi},\ and\ \citenamefont {Tsushima}}]{Sugano1971}%
  \BibitemOpen
  \bibfield  {author} {\bibinfo {author} {\bibfnamefont {S.}~\bibnamefont
  {Sugano}}, \bibinfo {author} {\bibfnamefont {K.}~\bibnamefont {Aoyagi}},\
  and\ \bibinfo {author} {\bibfnamefont {K.}~\bibnamefont {Tsushima}},\
  }\bibfield  {title} {\bibinfo {title} {Exciton absorption lines in
  antiferromagnetic rare-earth orthochromites –with particular reference to
  {YCrO}$_3^{–}$},\ }\href@noop {} {\bibfield  {journal} {\bibinfo  {journal}
  {J. Phys. Soc. Jap.}\ }\textbf {\bibinfo {volume} {31}},\ \bibinfo {pages}
  {706} (\bibinfo {year} {1971})}\BibitemShut {NoStop}%
\bibitem [{\citenamefont {Mero}\ \emph {et~al.}(2021)\citenamefont {Mero},
  \citenamefont {Lai}, \citenamefont {Du},\ and\ \citenamefont
  {Liu}}]{Rea:2021}%
  \BibitemOpen
  \bibfield  {author} {\bibinfo {author} {\bibfnamefont {R.~D.}\ \bibnamefont
  {Mero}}, \bibinfo {author} {\bibfnamefont {C.-H.}\ \bibnamefont {Lai}},
  \bibinfo {author} {\bibfnamefont {C.-H.}\ \bibnamefont {Du}},\ and\ \bibinfo
  {author} {\bibfnamefont {H.-L.}\ \bibnamefont {Liu}},\ }\bibfield  {title}
  {\bibinfo {title} {{Spectroscopic Signature of Spin-Charge-Lattice Coupling
  in CuB$_2$O$_4$}},\ }\href {https://doi.org/10.1021/acs.jpcc.1c00111}
  {\bibfield  {journal} {\bibinfo  {journal} {J. Phys. Chem. C}\ }\textbf
  {\bibinfo {volume} {125}},\ \bibinfo {pages} {4322–4329} (\bibinfo {year}
  {2021})}\BibitemShut {NoStop}%
\bibitem [{\citenamefont {Petrakovskii}\ \emph {et~al.}(2000)\citenamefont
  {Petrakovskii}, \citenamefont {Sablina}, \citenamefont {Velikanov},
  \citenamefont {Vorotynov}, \citenamefont {Volkov},\ and\ \citenamefont
  {Bovina}}]{Petrakovskii2000}%
  \BibitemOpen
  \bibfield  {author} {\bibinfo {author} {\bibfnamefont {G.~A.}\ \bibnamefont
  {Petrakovskii}}, \bibinfo {author} {\bibfnamefont {K.~A.}\ \bibnamefont
  {Sablina}}, \bibinfo {author} {\bibfnamefont {D.~A.}\ \bibnamefont
  {Velikanov}}, \bibinfo {author} {\bibfnamefont {A.}~\bibnamefont
  {Vorotynov}}, \bibinfo {author} {\bibfnamefont {N.~V.}\ \bibnamefont
  {Volkov}},\ and\ \bibinfo {author} {\bibfnamefont {A.~F.}\ \bibnamefont
  {Bovina}},\ }\bibfield  {title} {\bibinfo {title} {Synthesis and magnetic
  properties of copper metaborate single crystals, {CuB}$_2${O}$_4$},\
  }\href@noop {} {\bibfield  {journal} {\bibinfo  {journal} {Crystallogr.
  Rep.}\ }\textbf {\bibinfo {volume} {45}},\ \bibinfo {pages} {853} (\bibinfo
  {year} {2000})}\BibitemShut {NoStop}%
\bibitem [{\citenamefont {McClure}(1959)}]{MCCLUREbook}%
  \BibitemOpen
  \bibfield  {author} {\bibinfo {author} {\bibfnamefont {D.~S.}\ \bibnamefont
  {McClure}},\ }\href
  {https://doi.org/https://doi.org/10.1016/S0081-1947(08)60569-X} {\emph
  {\bibinfo {title} {Electronic Spectra of Molecules and Ions in Crystals Part
  II. Spectra of Ions in Crystals: Part II. Spectra of Ions in Crystals}}},\
  edited by\ \bibinfo {editor} {\bibfnamefont {F.}~\bibnamefont {Seitz}}\ and\
  \bibinfo {editor} {\bibfnamefont {D.}~\bibnamefont {Turnbull}},\ \bibinfo
  {series} {Solid State Physics}, Vol.~\bibinfo {volume} {9}\ (\bibinfo
  {publisher} {Academic Press},\ \bibinfo {year} {1959})\ pp.\ \bibinfo {pages}
  {399--525}\BibitemShut {NoStop}%
\bibitem [{\citenamefont {Eremin}(2019)}]{Eremin2019_2}%
  \BibitemOpen
  \bibfield  {author} {\bibinfo {author} {\bibfnamefont {M.~V.}\ \bibnamefont
  {Eremin}},\ }\bibfield  {title} {\bibinfo {title} {On the theory of
  magnetoelectric coupling in {LiCu}$_2${O}$_2$},\ }\href
  {https://doi.org/10.1134/S1063776119110037} {\bibfield  {journal} {\bibinfo
  {journal} {J. Exp. Theor. Phys.}\ }\textbf {\bibinfo {volume} {6}},\ \bibinfo
  {pages} {990} (\bibinfo {year} {2019})}\BibitemShut {NoStop}%
\bibitem [{\citenamefont {Abragam}\ and\ \citenamefont
  {Bleaney}(2012)}]{abragam2012electron}%
  \BibitemOpen
  \bibfield  {author} {\bibinfo {author} {\bibfnamefont {A.}~\bibnamefont
  {Abragam}}\ and\ \bibinfo {author} {\bibfnamefont {B.}~\bibnamefont
  {Bleaney}},\ }\href@noop {} {\emph {\bibinfo {title} {Electron Paramagnetic
  Resonance of Transition Ions}}}\ (\bibinfo  {publisher} {Oxford University
  Press, Oxford},\ \bibinfo {year} {2012})\BibitemShut {NoStop}%
\bibitem [{\citenamefont {Sugano}\ \emph {et~al.}(1970)\citenamefont {Sugano},
  \citenamefont {Tanabe},\ and\ \citenamefont
  {Kamimura}}]{SuganoTanabeKamimura}%
  \BibitemOpen
  \bibfield  {author} {\bibinfo {author} {\bibfnamefont {S.}~\bibnamefont
  {Sugano}}, \bibinfo {author} {\bibfnamefont {Y.}~\bibnamefont {Tanabe}},\
  and\ \bibinfo {author} {\bibfnamefont {H.}~\bibnamefont {Kamimura}},\
  }\href@noop {} {\emph {\bibinfo {title} {Multiplets of Transition-Metal Ions
  in Crystals}}}\ (\bibinfo  {publisher} {Academic Press, New York},\ \bibinfo
  {year} {1970})\BibitemShut {NoStop}%
\bibitem [{\citenamefont {Eremenko}\ and\ \citenamefont
  {Novikov}(1970)}]{Eremenko1970}%
  \BibitemOpen
  \bibfield  {author} {\bibinfo {author} {\bibfnamefont {M.~V.}\ \bibnamefont
  {Eremenko}}\ and\ \bibinfo {author} {\bibfnamefont {V.~P.}\ \bibnamefont
  {Novikov}},\ }\bibfield  {title} {\bibinfo {title} {Davydov splitting of the
  exciton line in antiferromagnetic {RbMnF}$_{3}$},\ }\href@noop {} {\bibfield
  {journal} {\bibinfo  {journal} {JETP Letters}\ }\textbf {\bibinfo {volume}
  {11}},\ \bibinfo {pages} {326} (\bibinfo {year} {1970})}\BibitemShut
  {NoStop}%
\bibitem [{\citenamefont {Eremenko}\ \emph {et~al.}(1992)\citenamefont
  {Eremenko}, \citenamefont {Litvinenko}, \citenamefont {Kharchenko},\ and\
  \citenamefont {Naumenko}}]{Eremenkobook}%
  \BibitemOpen
  \bibfield  {author} {\bibinfo {author} {\bibfnamefont {V.~V.}\ \bibnamefont
  {Eremenko}}, \bibinfo {author} {\bibfnamefont {Y.~G.}\ \bibnamefont
  {Litvinenko}}, \bibinfo {author} {\bibfnamefont {N.~K.}\ \bibnamefont
  {Kharchenko}},\ and\ \bibinfo {author} {\bibfnamefont {V.~M.}\ \bibnamefont
  {Naumenko}},\ }\href@noop {} {\emph {\bibinfo {title} {Magneto-optics and
  Spectroscopy of Antiferromagnets}}}\ (\bibinfo  {publisher} {Springer, New
  York},\ \bibinfo {year} {1992})\BibitemShut {NoStop}%
\bibitem [{\citenamefont {Imbusch}(1978)}]{Imbusch1978}%
  \BibitemOpen
  \bibfield  {author} {\bibinfo {author} {\bibfnamefont {G.~F.}\ \bibnamefont
  {Imbusch}},\ }\bibfield  {title} {\bibinfo {title} {Luminescence from solids
  with high concentrations of transition metal ions},\ }in\ \href@noop {}
  {\emph {\bibinfo {booktitle} {Luminescence of Inorganic Solids}}},\ \bibinfo
  {editor} {edited by\ \bibinfo {editor} {\bibfnamefont {B.~D.}\ \bibnamefont
  {Bartolo}}, \bibinfo {editor} {\bibfnamefont {V.}~\bibnamefont {Godberg}},\
  and\ \bibinfo {editor} {\bibfnamefont {D.}~\bibnamefont {Pacheco}}}\
  (\bibinfo  {publisher} {Springer, Boston},\ \bibinfo {year} {1978})\ pp.\
  \bibinfo {pages} {155--180}\BibitemShut {NoStop}%
\bibitem [{\citenamefont {Boehm}\ \emph {et~al.}(2002)\citenamefont {Boehm},
  \citenamefont {Martynov}, \citenamefont {Roessli}, \citenamefont
  {Petrakovskii},\ and\ \citenamefont {Kulda}}]{Boehm2002}%
  \BibitemOpen
  \bibfield  {author} {\bibinfo {author} {\bibfnamefont {M.}~\bibnamefont
  {Boehm}}, \bibinfo {author} {\bibfnamefont {S.}~\bibnamefont {Martynov}},
  \bibinfo {author} {\bibfnamefont {B.}~\bibnamefont {Roessli}}, \bibinfo
  {author} {\bibfnamefont {G.}~\bibnamefont {Petrakovskii}},\ and\ \bibinfo
  {author} {\bibfnamefont {J.}~\bibnamefont {Kulda}},\ }\href@noop {}
  {\bibfield  {journal} {\bibinfo  {journal} {J. Magn. Magn. Mater.}\ }\textbf
  {\bibinfo {volume} {250}},\ \bibinfo {pages} {313} (\bibinfo {year}
  {2002})}\BibitemShut {NoStop}%
\bibitem [{\citenamefont {Anderson}(1959)}]{Anderson1959}%
  \BibitemOpen
  \bibfield  {author} {\bibinfo {author} {\bibfnamefont {P.~W.}\ \bibnamefont
  {Anderson}},\ }\bibfield  {title} {\bibinfo {title} {New approach to the
  theory of superexchange interactions},\ }\href@noop {} {\bibfield  {journal}
  {\bibinfo  {journal} {Phys. Rev.}\ }\textbf {\bibinfo {volume} {115}},\
  \bibinfo {pages} {2} (\bibinfo {year} {1959})}\BibitemShut {NoStop}%
\end{thebibliography}
\end{document}